\documentclass[fleqn,usenatbib]{mnras}

\usepackage{newtxtext,newtxmath}

\pdfoutput=1

\usepackage{ulem}

\usepackage[T1]{fontenc}

\DeclareRobustCommand{\VAN}[3]{#2}
\let\VANthebibliography\thebibliography
\def\thebibliography{\DeclareRobustCommand{\VAN}[3]{##3}\VANthebibliography}

\usepackage{graphicx}	
\usepackage{amsmath}	
\usepackage{amssymb}	
\usepackage{scalerel}
\usepackage{ifpdf}

\setcitestyle{citesep={,}}
\setcitestyle{citeset={,}}

\newcommand{\bzepsilon}{\epsilon_{\scaleto{\mathrm{CTM}}{3pt}}}
\newcommand{\diracdelta}{\delta_{\scaleto{\mathrm{D}}{4pt}}}
\newcommand{\linearpow}{\mathrm{P}_{\scaleto{\mathrm{L}}{4.5pt}}}
\newcommand{\legendre}{\mathrm{P}_{\scaleto{l}{4.5pt}}}
\newcommand{\lingrowth}{D_{\scaleto{1}{4.5pt}}}

\title[CTM]{The Cosmological Trajectories Method: Modelling cosmic structure formation in the non-linear regime}

\author[F. C. Lane et al.]{
F. C. Lane,$^{1}$\thanks{E-mail: flane@roe.ac.uk}
A. N. Taylor,$^{1}$\thanks{E-mail: ant@roe.ac.uk}
and D. Sorini$^{1}$\thanks{E-mail: sorini@roe.ac.uk}
\\
$^{1}$Scottish Universities Physics Alliance, Institute for Astronomy, School of Physics and Astronomy, University of Edinburgh, Royal Observatory, 
\\
Blackford Hill, Edinburgh, EH9 3HJ, U.K.}

\date{Accepted XXX. Received YYY; in original form ZZZ}

\pubyear{2021}

\begin{document}
\label{firstpage}
\pagerange{\pageref{firstpage}--\pageref{lastpage}}
\maketitle

\begin{abstract}
We introduce a novel approach, the Cosmological Trajectories Method (CTM), to model nonlinear structure formation in the Universe by expanding gravitationally-induced particle trajectories around the Zel'dovich approximation. A new Beyond Zel'dovich  approximation is presented, which expands the CTM to leading second-order in the gravitational interaction and allows for post-Born gravitational scattering. In the Beyond Zel'dovich approximation we derive the exact expression for the matter clustering power spectrum. This is calculated to leading order and is available in the \textsc{CTM Module}. We compare the Beyond Zel’dovich approximation power spectrum and correlation function to other methods including 1-loop Standard Perturbation Theory (SPT), 1-loop Lagrangian Perturbation Theory (LPT), and Convolution Lagrangian Perturbation Theory (CLPT). We find that the Beyond Zel’dovich approximation power spectrum performs well, matching simulations to within $\pm{10}\%$, on mildly non-linear scales, and at redshifts above $z=1$ it outperforms the Zel'dovich approximation. We also find that the Beyond Zel'dovich approximation models the BAO peak in the correlation function at $z=0$ more accurately, to within $\pm{5}\%$ of simulations, than the Zel'dovich approximation, SPT 1-loop and CLPT.
\end{abstract}




\section{Introduction}
\label{sec: intro}

Deciphering how the cosmic web and large-scale structure is formed in our Universe is an essential part of understanding cosmology. Better knowledge of large-scale structure formation will allow us to extract more information from current~\citep[e.g.][]{planck2018,kids2017,des2017} and future observations of our Universe. Gathering more statistical information from current and upcoming surveys such as the Dark Energy Spectroscopic Instrument~\citep{desi}, the Vera Rubin Observatory~\citep{lsst} and \textit{Euclid}~\citep{euclid} will lead to tighter constraints on viable cosmological, gravity and structure formation models.

Modelling the cosmic web involves knowing how structures form under the influence of gravity. In the first approximation, the equations governing the evolution of density perturbations can be linearised. While this approach is accurate enough to describe the large-scale modes, it inevitably breaks down on small scales, where the local density field can become much larger than the average background density of the Universe. The breakdown of linear theory was found to occur around Fourier modes with wavenumber $k>0.1\ \mathrm{h}\ \mathrm{Mpc}^{-1}$~\citep{sugiyama2014a,mcquinn2016a}, hence this regime is generally referred to as ``non-linear regime''. A further degree of complexity comes into play when considering the impact of baryonic effects on galactic scales, such as winds ejected due to supernovae explosions or jets from active galactic nuclei \cite[see e.g.][for a review]{feedback_rev}.

Because of the non-linear and interconnected nature of the physical processes driving structure formation, the current preferred method for investigating structure formation in the non-linear regime is to run large cosmological simulations. N-body (dark matter only) and hydrodynamic (dark matter and baryons) simulations can be used to simulate the gravitational evolution of structure in the Universe. One of the first large N-body simulations, the Millennium simulation~\citep{Springel2005}, modelled the evolution of around a million dark matter particles from $z=127$ to $z=0$. Recent hydrodynamic simulations such as  EAGLE~\citep{Crain2015, Schaye2015}, Illustrius-TNG~\citep{weinberger2017,pillepich2018}, the New Horizon runs~\citep{newhorizon2011, newhorizon2020} and Simba~\citep{dave2019} have furthered our understanding of structure formation and baryonic effects. 

The large volume and high precision of data from forthcoming surveys demands at least comparable accuracy in theoretical models of structure formation. For this reason, simulations would need to probe a wide range of scales, while retaining high enough resolution to properly capture small-scale physics. However, the consequent computational cost in terms of memory and computer time hinders the exploration of a wide parameter space. This represents an issue when testing multiple theories of gravity and cosmological models, which typically requires obtaining predictions for several choices of the underlying parameters. Thus, there is clearly an interest for searching alternative and less costly methods.

Cosmological emulators provide a way of predicting the non-linear growth of structure for a range of cosmological parameters and some modified gravity theories. Emulators are generally trained on large sets of high-resolution simulation runs but once they have been trained on the simulation output they can be made publicly available for the community to utilise. In this paper, we will utilise the \textsc{Euclid Emulator}~\citep{euclidemu} which was developed for the \textit{Euclid} space telescope and was trained on a sample of 100 input runs of \textsc{PKDGRAV}3~\citep{Stadel2002,potter2016pkdgrav3}. Other examples of emulators include \textsc{CosmicEmu}~\citep{cosmicemu1, cosmicemu2, cosmicemu3, cosmicemu4, cosmicemu5} trained on the Coyote Universe simulations, \textsc{mgemu}~\citep{mgemu} an emulator that can model the ratio between the $\Lambda$-CDM power spectrum and the Hu-Sawiki $f\left(R\right)$ gravity~\citep{Hu2007} power spectrum, the \textsc{AEMULUS}~\citep{aemulus} project and the \textsc{Dark Quest}~\citep{darkquest} project. Although both simulations and emulators allow us to probe the non-linear regime accurately, analytic techniques can allow us to see how density field correlations arise more easily. 

One technique that does not require the running of simulations and can give a more in-depth insight into structure formation is perturbation theory. Standard Perturbation Theory (SPT) or Eulerian Perturbation Theory (EPT) can reproduce cosmological observations up to scales of around $k\sim{0.1}\ \mathrm{h}\ \mathrm{Mpc}^{-1}$, where non-linear gravitational and baryonic effects become important \citep[]{peebles1980,bertschinger1995,bouchet1996,bernardeau2002,carlson2009, bernardeau2013}. SPT can be extended to model smaller scales using IR-resummation (accounting for the physical effects of bulk flows) and loop corrections (next to leading order corrections), as demonstrated in \citet[]{crocce2006,crocce2006a,taruya2008,bernardeau2008,bernardeau2012,bernardeau2014} and \citet{blas2014}. Lagrangian Perturbation Theory (LPT) is another technique that can be used to model the formation of structure \citep[]{Moutarde1991,catelan1995,Buchert1992a,Buchert1993,bouchet1996,tatekawa2004,Rampf2012}. LPT, unlike SPT, follows the motion of particles through a system and has been found to be more accurate at equal orders than SPT \citep[]{matsubara2008,matsubara2008a,bouchet1995,carlson2013,white2014,catelan1995}. 

The Zel’dovich approximation~\citep[]{zeldovich1970, taylor1993, schneider1995,taylor1996,white2014}, a first-order LPT, is unique in that in 1D it is exact up until shell-crossing (the point at which streams of matter from different directions intersect) occurs. In 3D it behaves competitively with EPT and higher-order LPT. It is an intuitive method for describing how particles form the structures we see in the cosmic web~\citep{mcquinn2016a}.

As discussed in \citet{mcquinn2016a}, although techniques that aim to address the breakdown of perturbation theory on small scales have made an improvement (matching simulations up to $k=0.2\ \mathrm{h}\ \mathrm{Mpc}^{-1}$ as discussed in~\citealt{sugiyama2014a}), fundamental failings on these scales remain. For example, it is well known that when the overdensity field becomes large ($\delta\gg1$) these schemes are no longer valid. As mentioned above, both EPT and LPT also breakdown after shell-crossing occurs. However, we know that virialised structures in our Universe, such as dark matter haloes, are formed after shell-crossing occurs.

The Effective Field Theory of Large Scale Structure (EFTofLSS) is a method that aims to fix these issues by integrating out small wavelengths \citep[]{carrasco2012,carrasco2014,carrasco2014a,carroll2014,porto2014,senatore2015,vlah2015a,vlah2015,vlah2016a,vlah2016}, to reduce the uncontrolled small-scale perturbative effects affecting large scales. EFTofLSS generally requires either data from simulations or observations to fix free parameters in the theory. Other methods for extending perturbation theory into the non-linear regime including semi-classical propagators and methods based on field theory have been suggested in \citet{seljak2015,taruya2017,mcdonald2018,prokopec2017,friedrich2018,uhlemann2019,friedrich2019} and \citet{halle2020}.

A statistical mechanics approach to modelling gravitational interaction into the non-linear regime was introduced in \citet[]{bartelmann2014} and further developed in subsequent works \citep[]{bartelmann2014a,fabis2014,kozlikin2014, viermann2015,bartelmann2017,sorini2017,lilow2019,bartelmann2019a}. This theory is called Kinetic Field Theory (KFT). The theory was re-derived using particle trajectories in \citet{alihaimoud2015}. We will focus on the trajectories implementation of the technique. The initial results for the matter power spectrum \citep{bartelmann2014} hinted that the method could match current simulations. Another advantage of this method is that it has the potential to be easily adapted to multiple cosmological models, therefore allowing predictions to be made without running numerous simulations. 

In this paper, we will introduce the Cosmological Trajectories Method (CTM), which expands the trajectory around the Zel'dovich approximation. We present exact results for the matter power spectrum to leading second-order in the displacement field, in the Beyond Zel'dovich approximation, and show how an expanded version of the power spectrum can be calculated numerically in Section~\ref{sec: CTM}. Finally, in Section~\ref{subsec: compare BZ} we will compare the Beyond Zel'dovich approximation to other approximations including SPT 1-loop and LPT 1-loop. We find that the Beyond Zel'dovich approximation power spectrum matches the \textsc{Euclid Emulator}~\citep{euclidemu} more accurately than the Zel'dovich approximation above $z=1$. We also find that the Beyond Zel'dovich approximation captures the BAO peak in the two-point correlation function more accurately than SPT 1-loop, LPT 1-loop and CLPT at $z=0$. 

\section{Cosmological Trajectories Method (CTM)}
\label{sec: CTM}

The fundamental idea behind KFT~\citep{bartelmann2014, bartelmann2014a} is that an ensemble of dark matter particles moves dictated by some initial conditions until some redshift, $z_*$, when a gravitational interaction term is ``switched on'' as in an N-body simulation. This gravitational interaction term is then expanded perturbatively. This translates to an initial particle trajectory set by a Zel'dovich propagator, thus capturing the decaying velocity, with the addition of a gravitational correction term, the size of which is controlled by an expansion parameter, $\epsilon$. The formalism in KFT is based on field theory and therefore involves functional integrals, which is one motivation behind the work presented in~\citet{alihaimoud2015}.

In~\citet{alihaimoud2015}, the KFT results were re-derived in terms of particle trajectories and the Zel'dovich approximation. The  trajectory is found from the solution to the particle equations of motion,

\begin{align}
    \nonumber
    \mathbf{x}\left(\mathbf{q},z\right)=\mathbf{x}\left(\mathbf{q},z_*\right)&+\int^{z_*}_{z}\frac{dz'}{a'H\left(z'\right)}\mathbf{u}\left(\mathbf{q},z_*\right)
    \\
    \label{eqn: solution of eqns of motion}
    &-\epsilon\int^{z_*}_{z}\frac{dz'}{a'H\left(z'\right)}\int_{z'}^{z_*}\frac{dz''}{H\left(z''\right)}\nabla\phi\left(\mathbf{x}\left(\mathbf{q},z''\right),z''\right),
\end{align}

\noindent where $z_*$ is the redshift when the gravitational terms are ``switched on'', $z$ is redshift, $a$ is the scale factor, $H$ is the Hubble parameter, $\phi$ is the gravitational potential, $\mathbf{x}$ and $\mathbf{q}$ are the Eulerian and Lagrangian positions and $\mathbf{u}=a\mathbf{v}$ where $\mathbf{v}$ is the proper velocity. In~\citet{alihaimoud2015} the initial position $\mathbf{x}\left(\mathbf{q},z_*\right)$ and velocity $\mathbf{u}\left(\mathbf{q},z_*\right)$ at time $z_*$ are set by the Zel'dovich approximation. The gravitational interaction term, the final term in Equation (\ref{eqn: solution of eqns of motion}), is found via the Poisson equation and the overdensity field, calculated in the Zel'dovich approximation. The final result derived in~\citet{alihaimoud2015} is the power spectrum to first-order in this interaction term but is exact for the Zel'dovich density field.

The result and method presented in~\citet{alihaimoud2015} is very promising. However in both that approach, and in KFT, the correct linear growth is not recovered on large scales from the expansion of Equation (\ref{eqn: solution of eqns of motion}). This is because to the lowest-order gravitational interaction takes too long to overcome the damping effect of the expansion, leading to an underestimate of the displacement of particles and growth of structure. In both cases a re-normalisation of the predicted linear matter-density power spectrum is required.

This problem motivates us to introduce the Cosmological Trajectories Method (CTM), which expands the gravitational interaction around the free-field Zel'dovich approximation. This guarantees that to first order in the displacement we match linear theory, and to second-order we include the effects of gravitational scattering. This expansion avoids the non-local gravitational terms that appear at second-order in LPT~\citep{matsubara2008, matsubara2008a, Buchert1993}. One issue is that the free-field Zel'dovich approximation is already an approximation to gravitational collapse, so to avoid double-counting terms we remove a linear term from the gravitational interaction. A full derivation of the CTM is given in Appendix~\ref{appen gtc: general traj calc} and the CTM trajectory is presented in Equation~(\ref{appen gtc: general traj split up}).

In this paper, we will focus on the implementation where the gravitational interaction term in the CTM trajectory is expanded to second order in the displacement field. We shall refer to this as the Beyond Zel'dovich approximation, where the trajectory is given by (see Appendix~\ref{appen gtc: general traj calc} for a full derivation)

\begin{eqnarray}
    x_i\left(\mathbf{q},z\right)={q_i}  \!\!\!\!\!\! &+& \!\!\!\!\!\! A\left(z\right) \Psi_i\left(\mathbf{q},z_i\right)
     \nonumber \\
    &+& \!\!\!\!\!\!
    B_\epsilon \!\left(z\right)\Psi_j\left(\mathbf{q},z_i\right)
    \left({E}_{ij}(\mathbf{q},z_i) 
      +\frac{1}{3}\delta^{(0)}\!(\mathbf{q},z_i)\delta_{ij}\right) .
    \label{eqn: general trajectory full tidal}
\end{eqnarray}

\noindent Here $q_i$ is the initial Lagrangian position of the particles, $\Psi_i$ is the linear displacement field, $\delta^{\left(0\right)}$ is the linear overdensity field,

\begin{equation}
    E_{ij}= \left(\nabla_i\nabla_j\nabla^{-2}-\frac{1}{3}\delta_{ij}\right) \delta^{\left(0\right)},
\end{equation}

\noindent is a dimensionless,  trace-free linear tidal tensor, and $z_i$ is some initial  redshift. Equation~(\ref{eqn: general trajectory full tidal}) is the leading lowest-order gravitational correction to the Zel'dovich approximation, describing post-Born gravitational deflections from the unperturbed trajectory.

As we are expanding around the Zel'dovich approximation, the linear time-dependence function in Equation~(\ref{eqn: general trajectory full tidal}) is $A\left(z\right)=\frac{\lingrowth\left(z\right)}{\lingrowth\left(z_i\right)}$.  We note there is freedom to choose other time-dependencies, but this ensures the lowest-order theory matches linear growth on large scales. The second time-dependent function $B_\epsilon\left(z\right)$ in Equation~(\ref{eqn: general trajectory full tidal}) is derived in Appendix~\ref{appen gtc: general traj calc} to match the gravitational field, and is given by

\begin{equation}
    \label{eqn: B}
    B_\epsilon\left(z\right)=-\bzepsilon\omega_0^2\int_{z_i}^{z}\frac{dz'}{a' H\left(z'\right)}\int_{z_i}^{z'}\frac{dz''}{H\left(z''\right)}\left(\frac{\lingrowth\left(z''\right)}{\lingrowth\left(z_i\right)}\right)^2.
\end{equation}

\noindent  where $\bzepsilon$ controls the size of the higher-order gravitational term (the tidal tensor) and $\omega^2_0=H^2_0\Omega_m$. Note that in principle one could use the linear growth factor for a scale-independent modified gravity theory~\citep{Clifton2012,Nojiri2017} instead. Comparing the Beyond Zel'dovich approximation time dependence to that obtained in \citet{bartelmann2014, bartelmann2014a} and \citet{alihaimoud2015}, we see that as we do not include the initial, decaying velocity term. Instead the particle follows a Zel'dovich trajectory and the displacement field is therefore proportional to the linear growth factor. In Appendix~\ref{appen gtc: application to kft} more detail on the application of second-order CTM to KFT is given.

There are two free parameters in the second-order CTM trajectory; the initial redshift $z_i$ and the expansion parameter $\bzepsilon$. The expansion parameter, $\bzepsilon$, controls the size of the gravitational terms. If one considers $\bzepsilon$ as a perturbative parameter then by definition it should be small ($\bzepsilon\ll$1). However, this parameter can also be interpreted in a physical sense as controlling how large the non-linear structures being modelled are. We would expect that larger $\bzepsilon$ values will increase the impact the tidal field has on non-linear structure formation.

\subsection{Calculating 2-point statistics using the CTM}
\label{subsec: calc 2-pt stats}

In this section, we give details on the calculation of the matter-density power spectrum for the Beyond Zel'dovich trajectory. We find that, assuming Gaussian initial conditions, we can calculate an exact expression for the matter power spectrum in this approximation. In order to explore the numerical implementation of this result, we expand around the exact solution. Our numerical results are available using the \textsc{CTM Module}. We begin with the statistical properties of the linear fields.

\subsubsection{Covariance matrix and correlation functions}
\label{subsubsec: covariance matrix}

The linear displacement field, $\Psi_i$, the tidal field, $E_{ij}$, and the linear overdensity field, $\delta^{\left(0\right)}$, are correlated Gaussian fields at the initial redshift, $z_i$. As we shall show, we can  calculate the matter power spectrum in the Beyond Zel'dovich approximation using the statistics of Gaussian fields~\citep[]{bardeen1986, vandeweygaert1996, taylor2000}. As the fields are Gaussian, they are fully specified by their covariance matrix, $\mathbf{C}$, which contains the correlation of the fields with each other at two different Lagrangian points, $\mathbf{q}_1$ and $\mathbf{q}_2$, at the initial redshift. 

We define a vector of the fields at each position,

\begin{equation}
    \mathbf{X}=\left(\Psi_i(\mathbf{q}_1),\ \Psi_i(\mathbf{q}_2),\ E_i(\mathbf{q}_1),\ E_i(\mathbf{q}_2),\ \delta^{\left(0\right)}(\mathbf{q}_1),\ \delta^{\left(0\right)}(\mathbf{q}_2)\right),
\end{equation}

\noindent where $E_i={\rm vec}( \mathbf{E})$ is the 6-dimensional vectorisation of the distinct terms in the symmetric tidal tensor $E_{ij}$. The covariance matrix of the vector, $\mathbf{C}=\langle{\mathbf{X}\mathbf{X}^T}\rangle$, is given by

\begin{equation}
\label{eqn: covariance matrix}
\mathbf{C}=\left[\begin{array}{cccccc}
C_{\Psi_i^1\Psi_j^1} \!\!\! &  
C_{\Psi_i^2\Psi_j^1} \!\!\! & 
C_{E_{i}^1\Psi_{j}^1} \!\!\! & 
C_{E_{i}^2\Psi_{j}^1}\!\!\! & C_{\delta^{\left(0\right)}_1\Psi_i^1} \!\!\!  & C_{\delta^{\left(0\right)}_2\Psi_i^1}
\\
C_{\Psi_i^1\Psi_j^2} \!\!\! &  
C_{\Psi_i^2\Psi_j^2}\!\!\! & 
C_{E_{i}^1\Psi_{j}^2}\!\!\!  & C_{E_{i}^2\Psi_{j}^2}\!\!\!  & C_{\delta^{\left(0\right)}_2\Psi_i^1} \!\!\! & C_{\delta^{\left(0\right)}_2\Psi_i^2}
\\
C_{\Psi_i^1E_{j}^1} \!\!\! &  
C_{\Psi_i^2E_{j}^1} \!\!\! & C_{E_{i}^1E_{j}^1}\!\!\!  & C_{E_{i}^2E_{j}^1} \!\!\! & C_{\delta^{\left(0\right)}_1E_{i}^1}\!\!\!  & C_{\delta^{\left(0\right)}_2E_{i}^1}
\\
C_{\Psi_i^1E_{j}^2} \!\!\! &  C_{\Psi_i^2E_{j}^2} \!\!\! & C_{E_{i}^1E_{j}^2} \!\!\! & C_{E_{i}^2E_{j}^2}\!\!\!  & C_{\delta^{\left(0\right)}_1E_{i}^2}\!\!\! & C_{\delta^{\left(0\right)}_2E_{i}^2}
\\
C_{\Psi_i^1\delta^{\left(0\right)}_1} \!\!\! & C_{\Psi_i^2\delta^{\left(0\right)}_1} \!\!\! & C_{E_{i}^1\delta^{\left(0\right)}_1} \!\!\! & C_{E_{i}^2\delta^{\left(0\right)}_1}\!\!\! & C_{\delta^{\left(0\right)}_1\delta^{\left(0\right)}_1} \!\!\! & C_{\delta^{\left(0\right)}_2\delta^{\left(0\right)}_1}
\\
C_{\Psi_i^1\delta^{\left(0\right)}_2} \!\!\! & C_{\Psi_i^2\delta^{\left(0\right)}_2} \!\!\! & C_{E_{i}^1\delta^{\left(0\right)}_2}  \!\!\! & C_{E_{i}^2\delta^{\left(0\right)}_2} \!\!\! & C_{\delta^{\left(0\right)}_1\delta^{\left(0\right)}_2}\!\!\!  & C_{\delta^{\left(0\right)}_2\delta^{\left(0\right)}_2}
\end{array}\right],
\end{equation}

\noindent where the numerical `1' and `2' indicate the position. The correlators of the linear density and displacement fields are given by

\begin{subequations}
\label{eqn: defs of correlations}
\begin{align}
\label{eqn: defs of correlations: sigma_ij}
C_{\Psi_i^1\Psi_j^2}&=\langle\Psi_i\left(\mathbf{q}_1\right)\Psi_j\left(\mathbf{q}_2\right)\rangle=\sigma_{ij}\left(q\right),
\\
\label{eqn: defs of correlations: pi_i}
C_{\delta^{\left(0\right)}_1\Psi_i^1}&=\langle\delta^{\left(0\right)}\left(\mathbf{q}_1\right)\Psi_i\left(\mathbf{q}_2\right)\rangle=\Pi_i\left(q\right),
\\
\label{eqn: defs of correlations: sigma_0}
C_{\delta^{\left(0\right)}_1\delta^{\left(0\right)}_2}&=\langle\delta^{\left(0\right)}\left(\mathbf{q}_1\right)\delta^{\left(0\right)}\left(\mathbf{q}_2\right)\rangle=\xi_0\left(q\right),
\end{align}
\end{subequations}

\noindent where $q=\left|\mathbf{q}_2-\mathbf{q}_1\right|$ is the distance between points, while the correlations of the vectorised tidal field are

\begin{subequations}
\label{eqn: defs of correlations 2}
\begin{align}
\label{eqn: defs of correlations: phi_ijk}
C_{E_{i}^1\Psi_{j}^2}&=\langle{E}_{i}\left(\mathbf{q}_1\right)\Psi_j\left(\mathbf{q}_2\right)\rangle,
\\
\label{eqn: defs of correlations: eta_ijkl}
C_{E_{i}^1E_{j}^2}&=\langle{E}_{i}\left(\mathbf{q}_1\right){E}_{j}\left(\mathbf{q}_2\right)\rangle,
\\
\label{eqn: defs of correlations: Sigma_ij}
C_{E_{i}^1\delta^{\left(0\right)}_2}&=\langle{E}_{i}\left(\mathbf{q}_1\right)\delta^{\left(0\right)}\left(\mathbf{q}_2\right)\rangle.
\end{align}
\end{subequations}

\noindent These can be written in terms of the correlations of the tensor tidal field,
 \begin{subequations}
 \label{eqn: defs of correlations 3}
 \begin{align}
     \langle{E}_{ij}\left(\mathbf{q}_1\right)\Psi_k\left(\mathbf{q}_2\right)\rangle = \Phi_{ijk}\left(q\right), \\
     \langle{E}_{ij}\left(\mathbf{q}_1\right){E}_{kl}\left(\mathbf{q}_2\right)\rangle  =\eta_{ijkl}\left(q\right), \\
     \langle{E}_{ij}\left(\mathbf{q}_1\right)\delta^{\left(0\right)}\left(\mathbf{q}_2\right)\rangle=\Sigma_{ij}\left(q\right).
 \end{align}
 \end{subequations}
 \noindent The correlation functions given in Equations~(\ref{eqn: defs of correlations}) and Equations~(\ref{eqn: defs of correlations 3}) are defined in Appendix~\ref{appen: corr funcs}.

\subsubsection{The full power spectrum}
\label{subsubsec: ps CTM}

The matter-density power spectrum, $\mathrm{P}\left(k\right)$, is defined by the correlator of the Fourier modes of the density field,

\begin{equation}
    \langle \delta\left(\mathbf{k}_1\right)\delta\left(\mathbf{k}_2\right) \rangle = \left(2\pi\right)^3\mathrm{P}\left(k\right)\diracdelta\left(\mathbf{k}_1+\mathbf{k}_2\right),
\end{equation}

\noindent where the expectation value is calculated by an ensemble average. The Fourier transform of the overdensity field is given by

\begin{equation}
    \label{eqn: overden field dirac delta ft}
    \left(2\pi\right)^3\diracdelta\left(\mathbf{k}\right) + \delta\left(\mathbf{k}\right) = \int {d^3q\ } \mathrm{e}^{i\mathbf{k}\cdot\mathbf{x}(\mathbf{q},z)},
\end{equation}

\noindent where $\mathbf{x}$ is the trajectory defined in Equation~(\ref{eqn: general trajectory full tidal}). Therefore, the power spectrum for the second-order CTM trajectory is given by

\begin{equation}
\begin{aligned}
    \label{eqn: power spec general var}
    \mathrm{P}&\left(k,z\right)=\int{d^3q}\ \mathrm{e}^{i\mathbf{k}\cdot\mathbf{q}}
    \\
    &\times\left[\left\langle\mathrm{e}^{ik_i\Psi_j\left(\mathbf{q}_1,z_i\right)\left(A\left(z\right)\delta_{ij}+B_\epsilon\left(z\right)E_{ij}\left(\mathbf{q}_1,z_i\right)+\frac{1}{3}B_\epsilon\left(z\right)\delta^{\left(0\right)}\left(\mathbf{q}_1,z_i\right)\delta_{ij}\right)} \right.\right.
    \\
    &\left.\left.\mathrm{e}^{-ik_i\Psi_j\left(\mathbf{q}_2,z_i\right)\left(A\left(z\right)\delta_{ij}+B_\epsilon\left(z\right)E_{ij}\left(\mathbf{q}_2,z_i\right)+\frac{1}{3}B_\epsilon\left(z\right)\delta^{\left(0\right)}\left(\mathbf{q}_2,z_i\right)\delta_{ij}\right)}\right\rangle-1\right].
\end{aligned}
\end{equation}

\noindent We can simplify this by introducing a new vector,

\begin{equation}
    \mathbf{K}=A(z) \left(k_i,\ -k_i,\ 0,\ 0,\ 0,\ 0\right),
\end{equation} 

\noindent with the same dimensionality as $\mathbf{X}$. If we define a new matrix $\mathbf{M}$, with the same dimensionality as $\mathbf{C}$,

\begin{equation}
\label{eqn: matrix M}
\mathbf{M}=\frac{i}{3}B_\epsilon{k_i}\left[\begin{array}{cccccc}
0 & 0 & -3\delta_{jk} & 0 & -1 & 0
\\
0 & 0 & 0 & 3\delta_{jk} & 0 & 1
\\
-3\delta_{jk} & 0 & 0 & 0 & 0 & 0
\\
0 & 3\delta_{jk} & 0 & 0 & 0 & 0
\\
-1 & 0 & 0 & 0 & 0 & 0
\\
0 & 1 & 0 & 0 & 0 & 0
\end{array}\right],
\end{equation}

\noindent the ensemble average in Equation~(\ref{eqn: power spec general var}) can be rewritten in the multivariate Gaussian form

\begin{multline}
\label{eqn: multivariate gaussian simple}
    \langle\mathrm{e}^{ik_i\Psi_j\left(\mathbf{q}_1\right)\left(A\delta_{ij}+B_\epsilon{E}_{ij}\left(\mathbf{q}_1\right)+\frac{1}{3}B_{\epsilon}\delta^{\left(0\right)}\left(\mathbf{q}_1\right)\delta_{ij}\right)}
    \\
    \hspace{2cm}\times\mathrm{e}^{-ik_i\Psi_j\left(\mathbf{q}_2\right)\left(A\delta_{ij}+B_{\epsilon}E_{ij}\left(\mathbf{q}_2\right)+\frac{1}{3}B_{\epsilon}\delta^{\left(0\right)}\left(\mathbf{q}_2\right)\delta_{ij}\right)}\rangle
    \\
    =\frac{1}{\left(2\pi\right)^{10}}\int{d}^{20} \! X \, \left|\det{\mathbf{C}}\right|^{-1/2}\mathrm{e}^{-\frac{1}{2}\mathbf{X}^T\mathbf{C}^{-1}\mathbf{X}}\mathrm{e}^{i \mathbf{K}\cdot\mathbf{X}}\mathrm{e}^{-\frac{1}{2}\mathbf{X}^T \mathbf{M}\mathbf{X}}.
\end{multline}

\noindent Equation~(\ref{eqn: multivariate gaussian simple}) can be integrated, resulting in an exact expression for the matter power spectrum for the second-order CTM trajectory,
\begin{multline}
    \label{eqn: power spec general}
    \mathrm{P}\left(k,z\right)=\int{d^3q}\ \mathrm{e}^{i\mathbf{k}\cdot\mathbf{q}}\left[\left|\det{\left(1+\mathbf{MC}\right)}\right|^{-1/2}\mathrm{e}^{-\frac{1}{2}\mathbf{K}^T\mathbf{C}\left[1+\mathbf{MC}\right]^{-1}\mathbf{K}}-1\right].
\end{multline}

\noindent This expression is the main result of the paper.

\subsubsection{Expansion of the power spectrum}
\label{subsubsec: expand power}

While Equation (\ref{eqn: power spec general}) is exact, and the matrix manipulation can in principle be carried out numerically, the integration is highly oscillatory and can be numerically unstable. To explore the features of the Beyond Zel'dovich approximation we shall expand the solution in such a way as the take advantage of existing algorithms to treat the integration, and to compare to other methods.

The argument of the exponential in Equation~(\ref{eqn: power spec general}) can be expanded;

\begin{equation}
    \label{eqn: expand exponent}
    \mathbf{K}^T\mathbf{C}\left[1+\mathbf{MC}\right]^{-1}\mathbf{K}\approx\mathbf{K}^T\mathbf{C}\mathbf{K}-\mathbf{K}^T\mathbf{C}\mathbf{MC}\mathbf{K} \, .
\end{equation}

\noindent The first term here is

\begin{equation}
    \label{eqn: first exponent term}
    \begin{aligned}
        \mathbf{K}^T\mathbf{C}\mathbf{K} =2A^2\left(z\right)\left[k^2\sigma^2_{\psi}\left(z_i\right)-k_ik_j\sigma_{ij}\left(q,z_i\right)\right] \, ,
    \end{aligned}
\end{equation}

 \noindent where $\sigma_{ij}\left(0,z_i\right)=\sigma^2_{\psi} \delta_{ij}$, while the second term  vanishes. We can  expand the determinant in Equation~(\ref{eqn: power spec general}) as
 
\begin{equation}
    \det (1+\mathbf{M}\mathbf{C}) 
    =\exp \left(\mathrm{tr} \ln (1+\mathbf{MC})\right)
    \approx \exp \left(-\frac{1}{2} 
    \mathrm{tr} \,\mathbf{MCMC}\right)
\end{equation}

\noindent where $\mathrm{tr}\left(\mathbf{MC}\right)=0$ and,

\begin{align}
    \notag
    \mathrm{tr}\left(\mathbf{MCMC}\right)=\frac{4}{9}B_\epsilon^2k_ik_j
    \left[\Pi_i\left(q\right)\Pi_j\left(q\right)
    +6\sigma_{in}\left(q\right){\Sigma}_{nj}\left(q\right)
    \right.
    \\
    \label{eqn: trace in terms of barred correlations}
    +\xi_0\left(q\right)\sigma_{ij}\left(q\right)
    \left.-\xi_0(0)\sigma^2_{\psi}\delta_{ij}\right].
\end{align}

\noindent In this approximation the power spectrum is
 
\begin{equation}
    \label{eqn: power spec 2}
    \mathrm{P}\left(k,z\right)=\int{d^3q} \mathrm{e}^{i\mathbf{k}\cdot\mathbf{q}}
    \left[\mathrm{e}^{-\frac{1}{2}[\mathbf{K}^T\mathbf{C}
    \mathbf{K}-\mathrm{tr}(\mathbf{MCMC})]}-1
    \right].
\end{equation}

\noindent To lowest order this reduces to the Zel'dovich power spectrum~\citep[]{taylor1993, schneider1995,taylor1996}. Both of the terms in the exponential in equation (\ref{eqn: power spec 2}) have a factor $k_i k_j$, so the function is Gaussian. The covariance of this Gaussian is the differential  displacement covariance. Hence, we can interpret the extra term as the lowest-order correction to the displacement covariance matrix due to gravitational scattering.

\subsubsection{Numerically calculating the full expanded power spectrum}
\label{subsubsec: calc expanded power spec}

It is useful to define the correlation function $\bar{\Sigma}_{ij}\left(q\right) = 
\langle\bar{E}_{ij}\left(\mathbf{q}_1\right)\delta^{\left(0\right)}\left(\mathbf{q}_2\right)\rangle$, which can be related to the un-barred correlation function,

\begin{equation}
    \bar\Sigma_{ij}\left(q\right)={\Sigma}_{ij}\left(q\right)+\frac{1}{3}\xi_0\left(q\right)\delta_{ij}.
\end{equation}

\noindent The correlations $\sigma_{ij},\ \Pi_{i}$ and $\bar{\Sigma}_{ij}$ can be split into irreducible components~\citep{vlah2015a,catelan2000,crittenden2001}. The method used to split $\sigma_{ij}$, $\Pi_{i}$ and $\bar{\Sigma}_{ij}$ is shown in Appendix~\ref{appen: corr funcs example}. The correlations $\Pi_{i}$ and $\bar{\Sigma}_{ij}$ can be expanded as

\begin{align}
\label{eqn: split Sigma}
&\bar{\Sigma}_{ij}\left(q\right)=D\left(q\right)\delta_{ij}+F\left(q\right)\hat{q}_i\hat{q}_j,
\\
\label{eqn: split Pi}
    &\Pi_{i}\left(q\right)=G\left(q\right)\hat{q}_i.
\end{align}

\noindent and $D,\ F$ and $G$ are defined as

\begin{subequations}
    \label{eqn: D, F  and G definitions}
    \begin{align}
        \label{eqn: D, F and G definitions: D definition}
    D\left(q\right)&=\frac{1}{6\pi^2}\int^{\infty}_{0}{dk\ }\left[j_0\left(kq\right)+j_2\left(kq\right)\right]k^2\ \linearpow\left(k\right),
    \\
    \label{eqn: D, F and G definitions: F definition}
    F\left(q\right)&=-\frac{1}{2\pi^2}\int^{\infty}_{0}{dk\ }j_2\left(kq\right)k^2\ \linearpow\left(k\right),
    \\
    \label{eqn: D, F and G definitions: G definition}
    G\left(q\right)&=-\frac{1}{2\pi^2}\int_0^{\infty}dk\  j_1\left(kq\right){k\ }\linearpow\left(k\right).
    \end{align}
\end{subequations}

\noindent Finally, the correlation of the displacement field can be decomposed as,
\begin{equation}
    \sigma_{ij}=\left(\sigma^2_\psi-\frac{1}{2}X'\left(q\right)\right)\delta_{ij}-\frac{1}{2}Y'\left(q\right)\hat{q}_i\hat{q}_j,
\end{equation}

\noindent with 

\begin{equation}
    \label{eqn: defs of X and Y}
    \begin{aligned}
    X'\left(q\right)&=\frac{1}{2\pi^2}\int^{\infty}_{0}{dk\ }\left[\frac{2}{3}-2\frac{j_1\left(kq\right)}{kq}\right]\linearpow\left(k,z_i\right)
    \\
    Y'\left(q\right)&=\frac{1}{2\pi^2}\int^{\infty}_{0}{dk\ }\left[6\frac{j_1\left(kq\right)}{kq}-2j_0\left(kq\right)\right]\linearpow\left(k,z_i\right).
    \end{aligned}
\end{equation}

\noindent Substituting the decomposed correlations into Equation~(\ref{eqn: trace in terms of barred correlations}) then splitting the integral into $k^2$ and $k^2\mu^2$ parts using the method for numerically calculating the Zel'dovich power spectrum described in \citet[]{schneider1995,carlson2013,sugiyama2014a} and \citet{vlah2015a}. We can use the kth moment of the integral to calculate the angular integral

\begin{equation}
    \label{eqn: mu squared integral}
    I_{k} = \int^{1}_{-1}{d\mu\ }\mu^k \mathrm{e}^{ia\mu}\mathrm{e}^{b^2\mu^2}
\end{equation}

\noindent which can be solved using the general prescription~\citep{schneider1995,vlah2015a},

\begin{equation}
    \label{eqn: general formula mu to k}
I_k=2\left(-i\right)^k\mathrm{e}^b\sum^{\infty}_{n=0}\left(-2b\right)^n\left(\frac{d}{dk}\right)^ka^{-n}j_n\left(a\right),
\end{equation}

\noindent where $j_n$ is a spherical Bessel function. The angular parts of Equation~(\ref{eqn: power spec general}) are calculated using an identity resulting in the power spectrum becoming,

\begin{multline}
    \label{eqn: full power with zero lag}
    \mathrm{P}\left(k,z\right)\approx2\pi\int^{\infty}_{0}{dq \ q^2}\int^{1}_{-1}{d\mu\ }\mathrm{e}^{ikq\mu}
    \\
    \left[\mathrm{e}^{-\frac{1}{2}k^2A^2\left(X'+\mu^2Y'\right)}\mathrm{e}^{B_\epsilon^2k^2\left(W'+\mu^2Z'\right)}-\mathrm{e}^{-k^2\sigma^2_{\psi}\left(A^2+\frac{1}{3}B_\epsilon^2\eta^2_E\right)}\right],
\end{multline}

\noindent where the second exponential term is a Dirac delta function at the origin and has been added to cancel oscillations as described in \citet{schneider1995}. The Beyond Zel'dovich power spectrum to second-order is finally given by

\begin{align}
    \notag
    \mathrm{P}\left(k,z\right)&=4\pi\int_0^\infty{dq\ q^2}\mathrm{e}^{-\frac{1}{2}k^2\left[A^2\left(X'+Y'\right)-2B_\epsilon^2\left(W'+Z'\right)\right]}
    \\
    \label{eqn: final traj power spec}
    &\times\sum_{n=0}^{\infty}\left[\frac{k\left(A^2Y'-2B_\epsilon^2Z'\right)}{q}\right]^nj_n\left(kq\right)
\end{align}

\noindent where 

\begin{subequations}
    \label{eqn: W and Z definitions}
    \begin{align}
        \label{eqn: W and Z definitions: W}
        W'&=-\frac{1}{3}\sigma^2_{\psi}\eta^2_E+\frac{1}{3}\left(\sigma^2_{\psi}-\frac{1}{2}X'\right)\left(2D-\frac{1}{3}\xi^2_0\right),
        \\
        \label{eqn: W and Z definitions: Z}
        Z'&=\frac{1}{9}G^2+\frac{2}{3}\left(\sigma^2_\psi-\frac{1}{2}X'\right)F-\frac{1}{3}Y'\left(D+F-\frac{1}{6}\xi^2_0\right).
    \end{align}
\end{subequations}

\noindent In the above expressions for $W'$ and $Z'$ all functions apart from $\sigma^2_\psi$ and $\eta^2_E$ are evaluated at $q$.

\section{The Beyond Zel'dovich Approximation}
\label{sec: the BZ approx}

The power spectra presented in the remainder of this paper have been calculated using the \textsc{CTM Module}\footnote{https://github.com/franlane94/CTM}. The initial power spectra and cosmological parameters are calculated using \textsc{classylss}\footnote{https://classylss.readthedocs.io/en/stable/} and the spherical Bessel integrals are calculated using \textsc{mcfit}\footnote{https://github.com/eelregit/mcfit}. The power spectra are calculated using \textit{Planck18}~\citep{planck2018} cosmology ($\Omega_m=0.3123,\ h=0.6737,\ n_s=0.9665$ and $\sigma_8=0.8102$). We also sum over $n=32$ spherical-Bessel functions when calculating the power spectra. See Appendix~\ref{appen ni: comments on ni} for more details on the numerical integration tools used in the \textsc{CTM Module}.

There are two free parameters in the second-order CTM trajectory: the initial redshift, $z_i$, and the expansion parameter, $\bzepsilon$. The choice of the initial redshift does not make a noticeable difference to the final power spectrum unless a very low value such as $z_i=10$ is chosen. Since we assume that the fields are initially Gaussian a sufficiently high value of $z_i$ must be chosen to not invalidate the method. In this paper, we will set $z_i=100$.

\begin{figure*}
 \includegraphics[width=\textwidth]{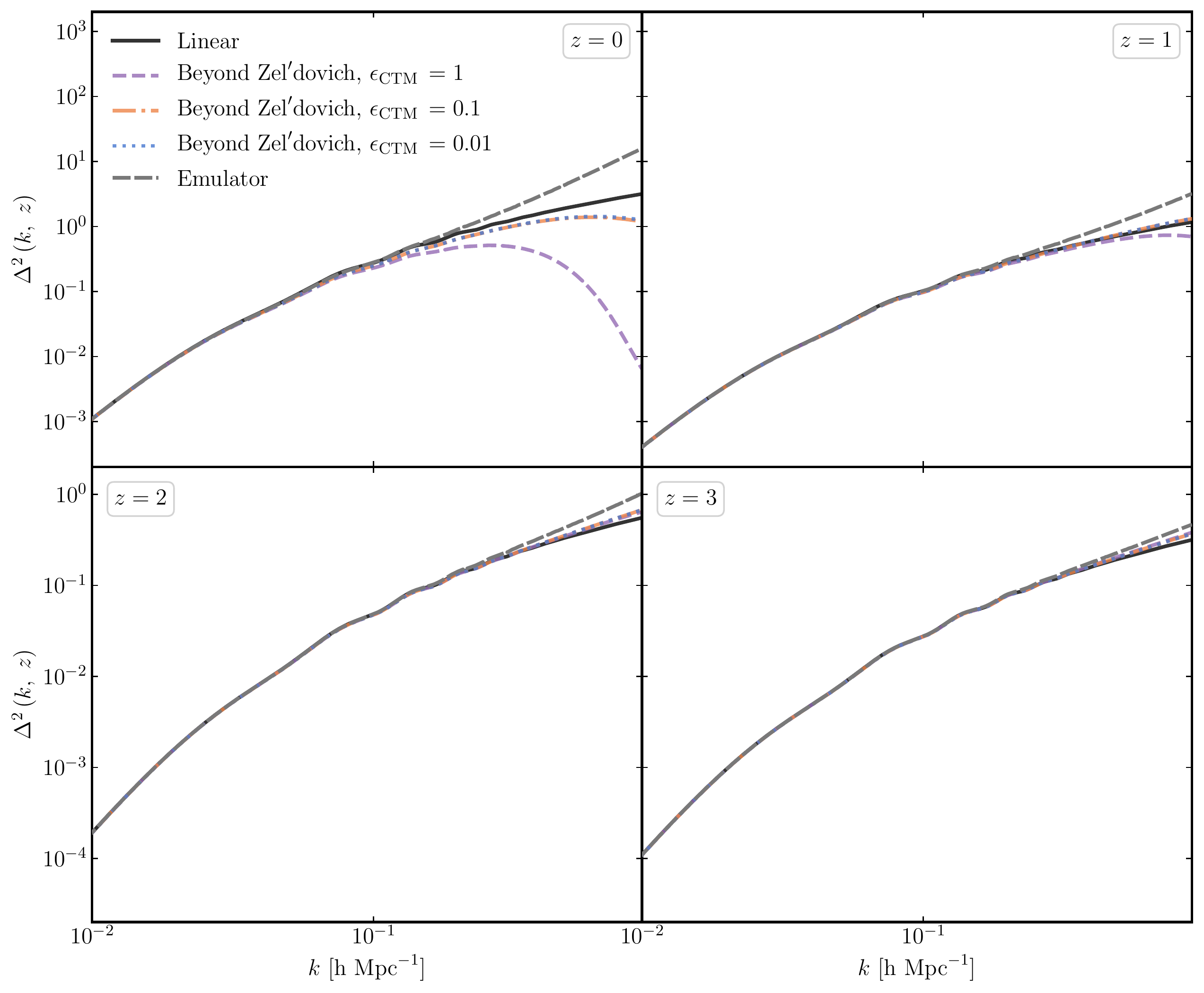}
 \caption{The dimensionless power spectrum for linear theory (black solid line), Beyond Zel'dovich with $\bzepsilon=1$ (dashed purple line), Beyond Zel'dovich with $\bzepsilon=0.1$ (dashed-dot orange line) and Beyond Zel'dovich with $\bzepsilon=0.01$ (dotted blue line) at $z=0$ in the top-left panel, at $z=1$ top-right panel, at $z=2$ in the lower-left panel and at $z=3$ in the lower-right panel. The \textsc{Euclid Emulator} result is shown in by grey dashed line in at the relevant redshift all panels.}
\label{fig:breakdown}
\end{figure*}

Recall that, the Beyond Zel'dovich approximation is the second-order CTM trajectory where the linear displacement terms are proportional to the linear growth factor. In Figure~\ref{fig:breakdown}, the dimensionless Beyond Zel'dovich power spectrum calculated with three $\bzepsilon$ values, $\bzepsilon=0.01,\ 0.1,\ 1$ at $z=0$ is shown in the top-left panel, at $z=1$ in the top-right panel, at $z=2$ in the lower-left panel and at $z=3$ in the lower-right panel. Results from the \textsc{Euclid Emulator}\footnote{https://github.com/miknab/EuclidEmulator/wiki/III)-Usage}~\citep{euclidemu} are shown in dashed grey lines. In~\citet{euclidemu}, the emulator is found to be $\pm 1\%$ accurate compared to simulations at $z=0$ and $\pm 1\%$ up to $k=1\ \mathrm{h}\ \mathrm{Mpc}^{-1}$ at $z=1$. Above $z=1$ it is around $\approx 3\%$ accurate. The emulator was built on a sample of 100 input runs of \textsc{PKDGRAV}3~\citep{Stadel2002,potter2016pkdgrav3}.

The power spectra shown in Figure~\ref{fig:breakdown} have been truncated at $k=0.9\ \mathrm{h}\ \mathrm{Mpc}^{-1}$. The second-order CTM trajectory is only applicable until this $k$-value as the method suffers from numerical issues beyond this point and it is difficult to disentangle these from physical effects. This is addressed in Appendix~\ref{appen ni: comments on ni}. At all redshifts, the Beyond Zel'dovich approximation calculated with $\bzepsilon=0.1$ and $\bzepsilon=0.01$ appears to have little effect on the trajectory. This was to be expected as the $\bzepsilon$ parameter controls the size of the gravitational correction to the Zel'dovich trajectory. The Beyond Zel'dovich approximation with $\bzepsilon=1$ performs well at redshifts above $z=2$ compared to the \textsc{Euclid Emulator}. However, at low redshifts the power spectrum is not boosted on small-scales, rather it is excessively damped. This is most likely due to shell-crossing and the particle trajectories overshooting on small-scales. We will present a solution to this excessive damping in Section~\ref{subsec: gauss damped init ps}. 

\begin{figure}
 \includegraphics[width=\columnwidth]{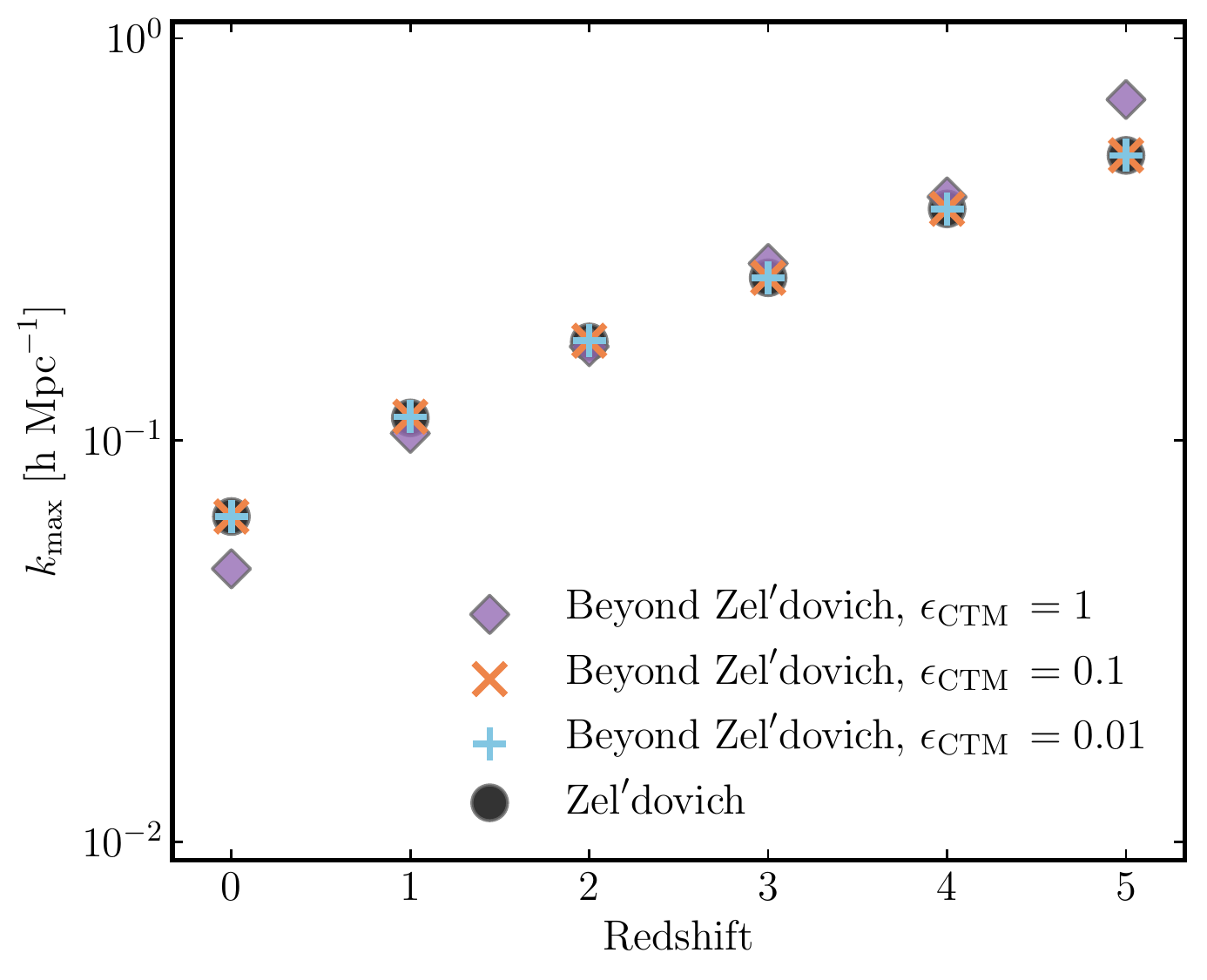}
 \caption{The maximum $k$-value reached, $k_\mathrm{max}$ , before the difference between the Beyond Zel'dovich approximation and the \textsc{Euclid Emulator} exceeds $\pm5\%$ versus redshift. The purple diamonds represent the Beyond Zel'dovich approximation calculated with $\bzepsilon=1$, the orange crosses are calculated with $\bzepsilon=0.1$ and the blue plus signs with $\bzepsilon=0.01$. The black circles represent the Zel'dovich approximation.}
 \label{fig:choice of epsilon}
\end{figure}

\vspace{-0.07cm}
The effect of the $\bzepsilon$ parameter on the Beyond Zel'dovich power spectrum is shown in more detail in Figure~\ref{fig:choice of epsilon}. The maximum $k$-value reached before the difference,

\begin{equation}
    \label{eqn: percent diff}
    \Delta_\mathrm{diff}=\frac{\mathrm{P}_\mathrm{calc}\left(k\right)-\mathrm{P}_\mathrm{emu}\left(k\right)}{\mathrm{P}_\mathrm{emu}\left(k\right)},
\end{equation}

\noindent where $\mathrm{P}_\mathrm{emu}$ is the power spectrum obtained using the \textsc{Euclid Emulator} exceeds $\Delta_\mathrm{diff}=\pm{0.05}$ is shown. 

One can see more clearly that the Beyond Zel'dovich approximation calculated with smaller values of $\bzepsilon$ (shown in blue plus signs and orange crosses) converges to the Zel'dovich approximation (shown in black circles). This validates our approximation as if $\bzepsilon=0$ the Beyond Zel'dovich approximation reduces to the Zel'dovich approximation. To have any improvement over the Zel'dovich approximation at high redshifts, we must have a value of $\bzepsilon=1$ (shown in purple diamonds). The Beyond Zel'dovich results in the remainder of this paper will be calculated with this value.

\subsection{A Gaussian damped initial power spectrum}
\label{subsec: gauss damped init ps}

To reduce the impact of the small-scale breakdown on larger scales, we will introduce a Gaussian damped initial power spectrum defined as
\begin{equation}
    \label{eqn: gauss damp ps}
    \mathrm{P}_\mathrm{damped}\left(k,z\right)=\mathrm{e}^{-\left(\frac{k}{k_c}\right)^2}\linearpow\left(k,z\right)
\end{equation}
\noindent where $k_c$ is the cut-off scale. 

\begin{figure}
 \includegraphics[width=\columnwidth]{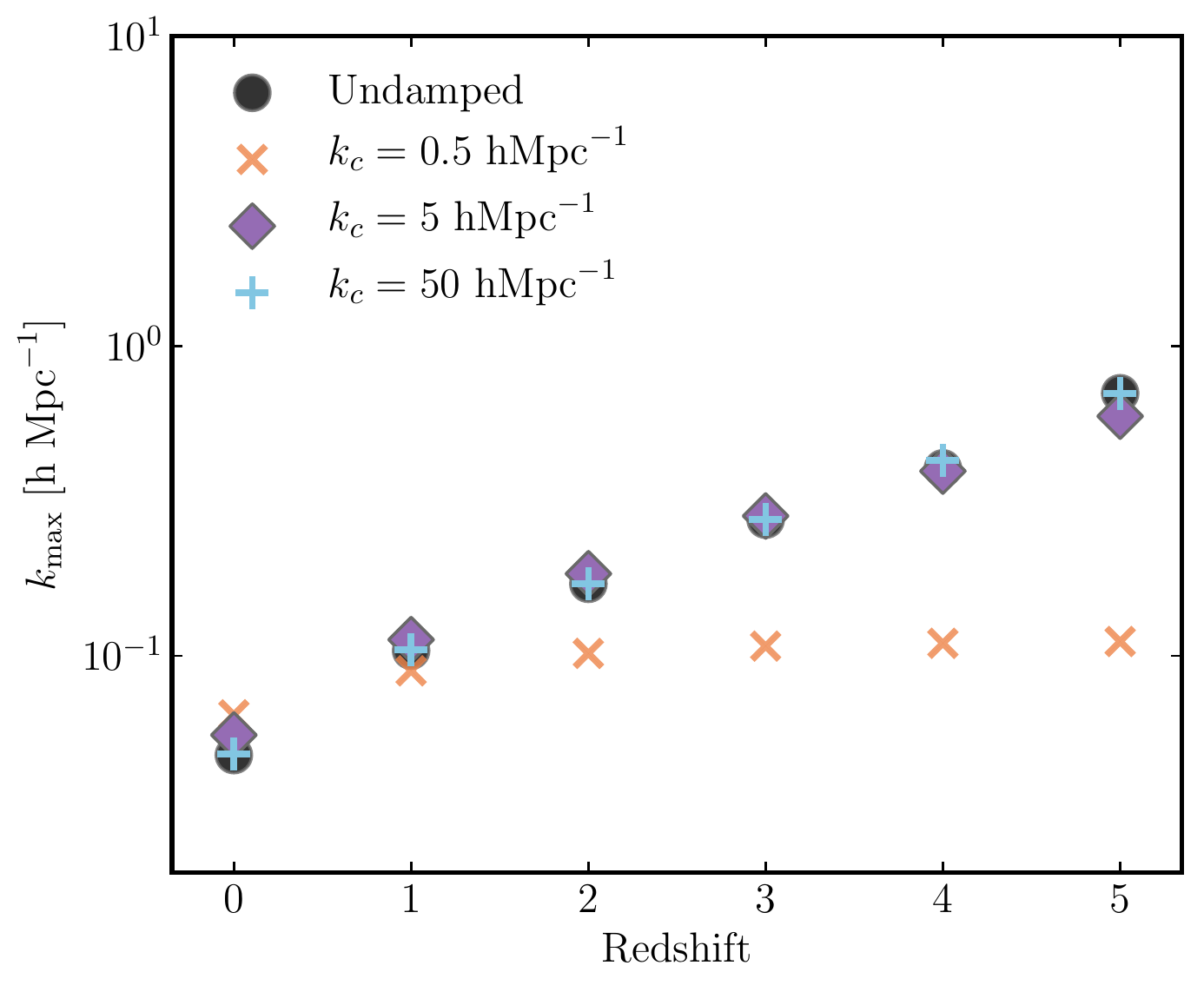}
 \caption{The maximum $k-$value reached before the difference between the Beyond Zel'dovich approximation power spectrum, calculated with $\bzepsilon=1$ and an initial Gaussian damped power spectrum shown in Equation~(\ref{eqn: gauss damp ps}), becomes larger than $\pm{5}\%$ is shown vs. redshift.}
 \label{fig: kc unzoomed}
\end{figure}

In Figure~\ref{fig: kc unzoomed}, the maximum $k-$value reached before the difference between the calculated Beyond Zel'dovich power spectrum and the emulator power spectrum becomes larger than $\Delta_\mathrm{diff}=\pm{5}\%$ is shown versus redshift. The Beyond Zel'dovich power spectra were calculated using $\bzepsilon=1$ and an initial Gaussian damped power spectrum with $k_c=50\ \mathrm{h}\ \mathrm{Mpc}^{-1}$ (blue plus signs), $k_c=5\ \mathrm{h}\ \mathrm{Mpc}^{-1}$ (purple diamonds) and $k_c=0.5\ \mathrm{h}\ \mathrm{Mpc}^{-1}$ (orange crosses). The highest cut-off value of $k_c=50\ \mathrm{h}\ \mathrm{Mpc}^{-1}$ has no noticeable effect and the lowest cut-off value of $k_c=0.5\ \mathrm{h\ }\mathrm{Mpc}^{-1}$ does not counteract the over damping on small scales until $z=0$. The largest and smallest cut-off values are too stringent and either restrict structure formation too much or too little in the desired regime. The value of $k_c=5\ \mathrm{h}\ \mathrm{Mpc}^{-1}$, however, appears to effectively remove the influence of the breakdown on larger scales at a wide range of redshifts. 

\begin{figure}
 \includegraphics[width=\columnwidth]{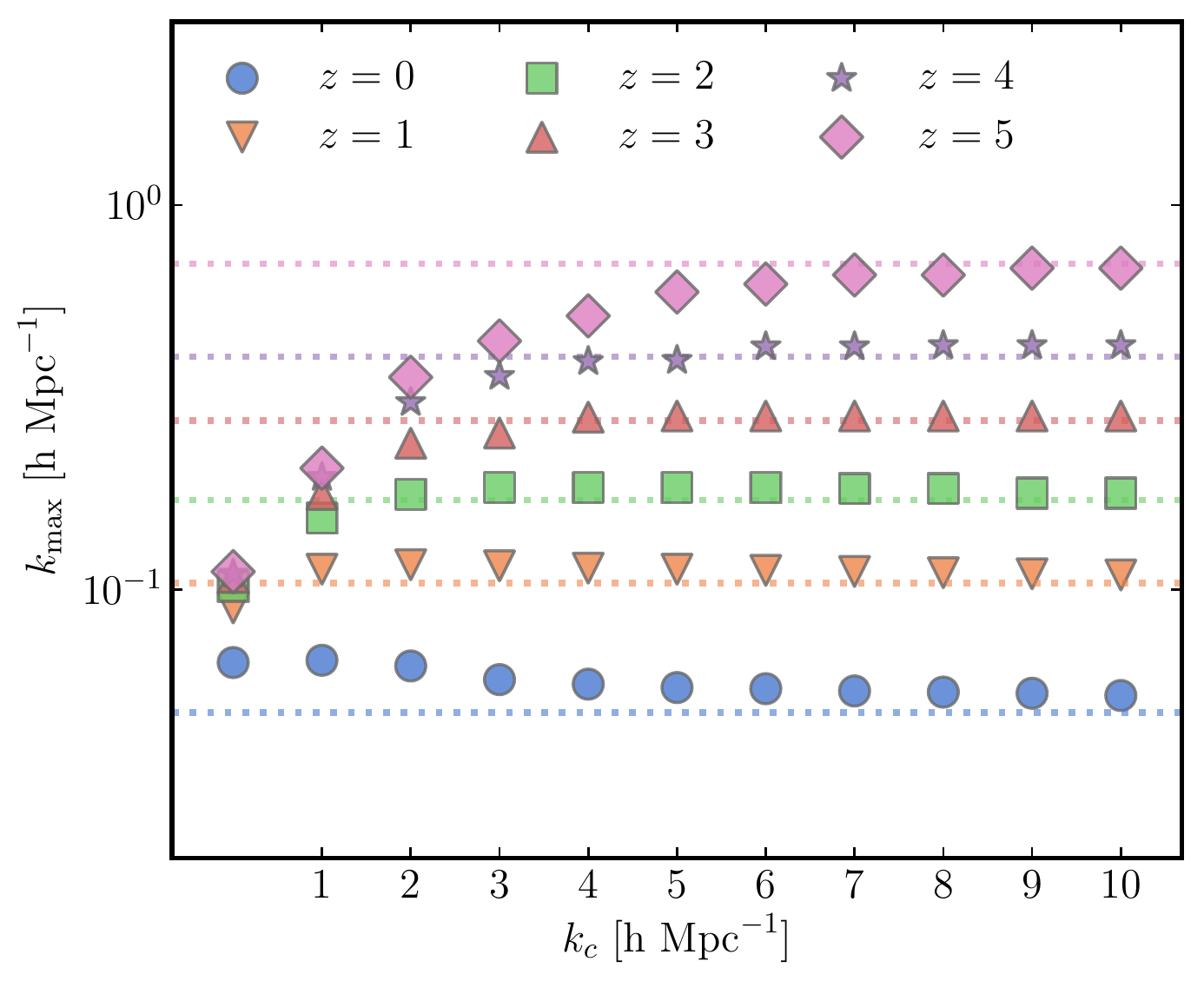}
 \caption{The maximum $k-$value reached before the difference between the Beyond Zel'dovich approximation power spectrum, calculated with $\bzepsilon=1$ and an initial Gaussian damped power spectrum in Equation~(\ref{eqn: gauss damp ps}), becomes larger than $\pm{5}\%$ is shown vs. the cut-off value $k_c$.}
 \label{fig: kc zoomed}
\end{figure}

Therefore, in Figure~\ref{fig: kc zoomed} a range of cut-off values centered around $k_c=5\ \mathrm{h}\ \mathrm{Mpc}^{-1}$ are tested. As was the case previously small cut-off values ($k_c<4\ \mathrm{h}\ \mathrm{Mpc}^{-1}$) have a detrimental effect on structure formation at high redshifts. We will choose to set $k_c=6\ \mathrm{h}\ \mathrm{Mpc}^{-1}$ to remove the effect of the breakdown on the Beyond Zel'dovich power spectrum for as wide a redshift range as possible. Hence, in all future plots the Beyond Zel'dovich power spectra are calculated with an initial Gaussian damped power spectrum with $k_c=6\ \mathrm{h}\ \mathrm{Mpc}^{-1}$.

Although not shown in this paper, we tested the dependence of the cut-off parameter $k_c$ on the cosmology chosen. We compared the value of $k_\mathrm{max}$ reached when the damped power spectrum was calculated using a value of $\Omega_m=0.3123$ and a value of $\Omega_m=0.3155$ (note all other cosmological parameters were kept the same). We found that there was a small difference of $\sim 5\%$ on average, implying that the cut-off scale is likely only weakly dependent on both cosmology and redshift. As we were comparing our results to the \textsc{Euclid Emulator} we were limited in the range of $\Omega_m$ values we could choose. We leave it to future work to obtain simulation data for a wider range of cosmological parameters to more stringently test the dependence of the cut-off on cosmology.

\subsection{Comparing the Beyond Zel'dovich approximation correlation function to other methods}
\label{subsec: compare bz corr func}

\begin{figure*}
 \includegraphics[width=\textwidth]{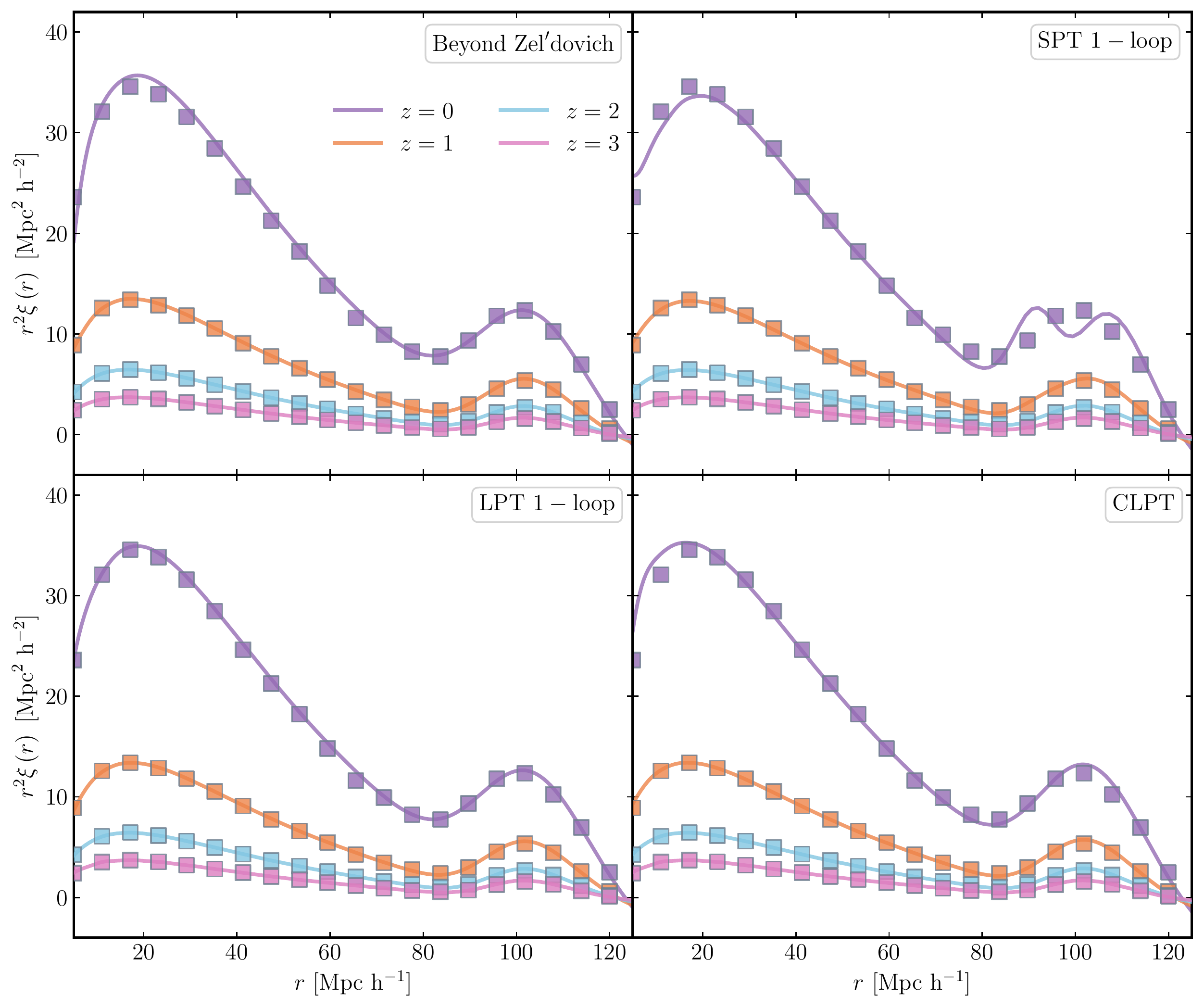}
 \caption{The scaled two-point correlation function calculated using the Beyond Zel'dovich approximation (upper-left panel), SPT 1-loop (upper-right panel), LPT 1-loop (lower-left panel) and CLPT (lower-right panel) is shown for four redshifts ($z=0$ in purple, $z=1$ in orange, $z=2$ in blue, $z=3$ in pink). The results from the \textsc{Euclid Emulator} are shown in squares.}
 \label{fig:corr compare}
\end{figure*}

In this section, we will investigate the performance of the Beyond Zel'dovich approximation for modelling the two-point correlation function. Specifically, we are interested in modelling the mildly non-linear regime, as this is where we find the BAO signal ($r\approx{100}\ \mathrm{Mpc}\ \mathrm{h}^{-1}$ or $k\approx{0.01}\ \mathrm{h}\ \mathrm{Mpc}^{-1}$) first detected in~\citet{Eisenstein2005} and~\citet{Cole2005}. In Figure~\ref{fig:corr compare}, the solid lines show the scaled correlation functions for the Beyond Zel'dovich approximation (upper-left panel), SPT 1-loop (upper-right panel) as discussed in \citet[]{crocce2006,crocce2006a,taruya2008,bernardeau2008,bernardeau2012,bernardeau2014, blas2014}, LPT 1-loop (lower-left panel) as described in~\citet{matsubara2008, matsubara2008a, carlson2009, vlah2015a, sugiyama2014a, carlson2013, mcquinn2016a, vlah2015a} and CLPT (lower-right panels) as described in \citet{carlson2013,wang2013, vlah2015a, vlah2015, vlah2016}. The different colours represent redshifts $z=0$ to $z=3$ and the squares show the results from the \textsc{Euclid Emulator}. In all comparison figures, the solid lines represent the Beyond Zel'dovich approximation.

One can see that above $z=1$ all perturbative schemes and approximations match the emulator results well. This is to be expected as the non-linear effects that cause the spatial deformation of the BAO peak are small.  In SPT 1-loop, the BAO peak is expected to grow in amplitude over time. This is true, however, the peak is expected to remain the same spatially. This is not accurate as bulk flows disrupt the shape of the peak~\citep{mcquinn2016a}. This is why in the upper right panel SPT 1-loop does not capture the BAO peak. 

The Beyond Zel'dovich approximation (upper-left panel), LPT 1-loop (lower-left panel), CLPT (lower-right panel), on the other hand, capture both the BAO peak and the small scales well. Both CLPT and the Beyond Zel'dovich approximation appear to be marginally less accurate on small scales than LPT 1-loop. Again, the performance of these methods is as expected. Methods based on the Zel'dovich approximation model the spatial evolution of the BAO peak more accurately due to their more precise modelling of spatial deformation. 

\begin{figure}
 \includegraphics[width=\columnwidth]{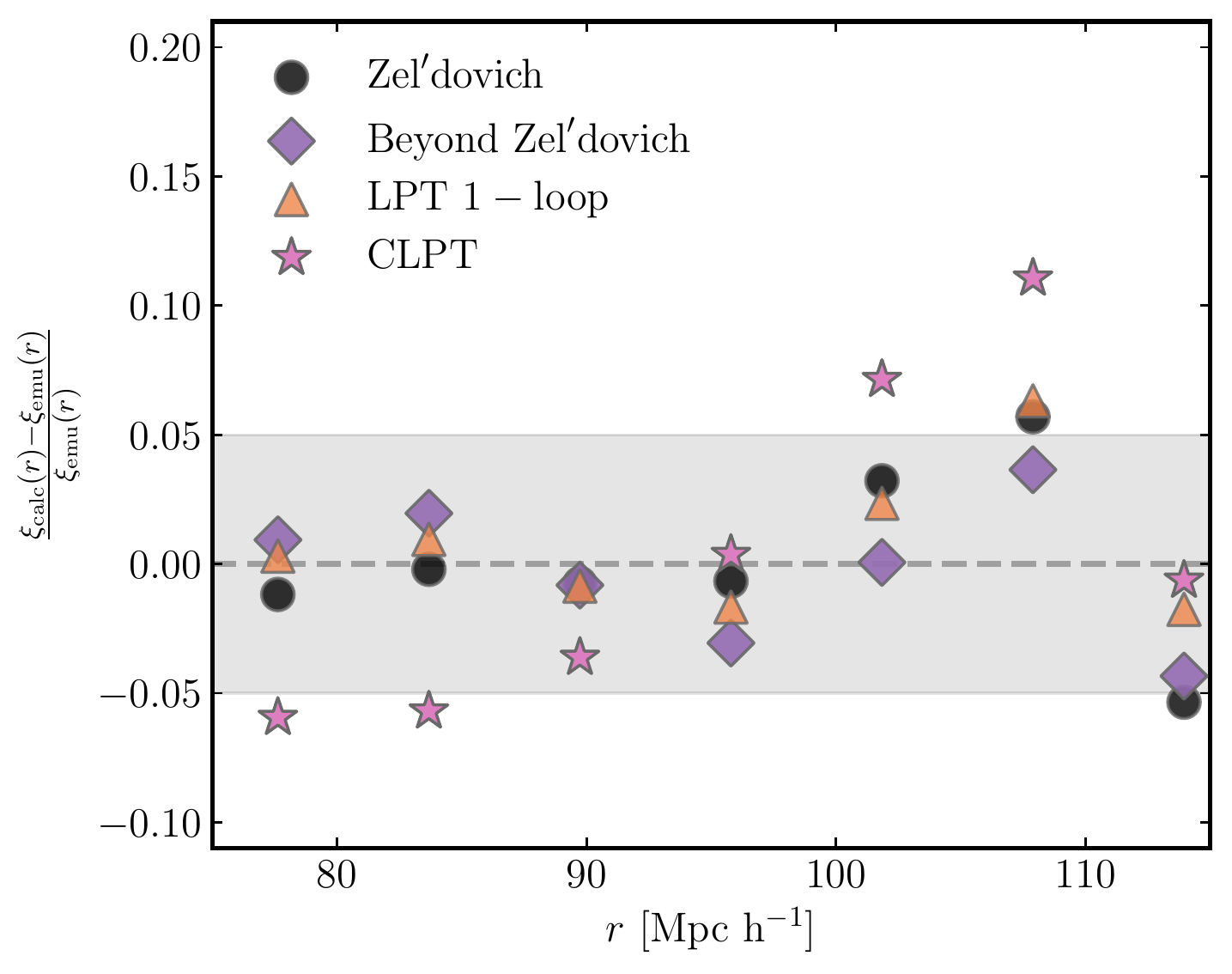}
 \caption{The difference, $\Delta_\mathrm{diff}$, between the Beyond Zel'dovich (purple diamonds), Zel'dovich approximation (black circles), LPT 1-loop (orange triangles) and CLPT (blue stars) correlation functions and the \textsc{Euclid Emulator} at $z=0$. The grey shaded region shows $\Delta_\mathrm{diff}\pm{0.05}$.}
 \label{fig:corr percent diff}
\end{figure}

In Figure~\ref{fig:corr percent diff}, the difference between the Beyond Zel'dovich, Zel'dovich approximation, LPT 1-loop and CLPT correlation functions and the emulator results are shown at $z=0$. The Beyond Zel'dovich approximation matches the \textsc{Euclid Emulator} more closely in the mildly non-linear regime than CLPT. The accuracy of the Beyond Zel'dovich approximation in the BAO peak regime is due to the inclusion of the tidal field term in Equation~(\ref{eqn: general trajectory full tidal}), as the spatial deformation responsible for the non-linear evolution of the BAO peak, is encoded within the tidal tensor. 

\subsection{Comparing the Beyond Zel'dovich approximation power spectrum to other methods}
\label{subsec: compare BZ}

\begin{figure*}
 \includegraphics[width=\textwidth]{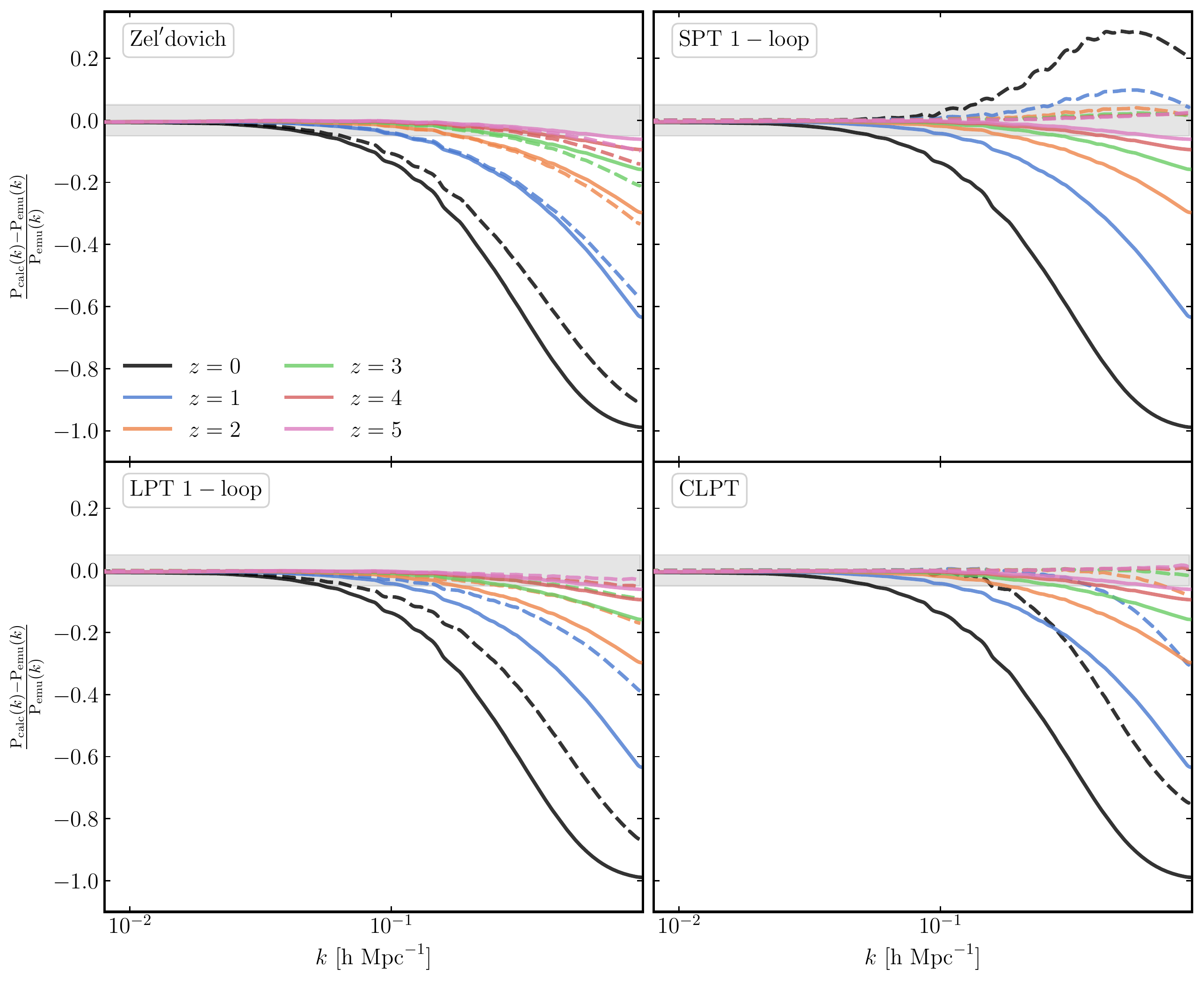}
 \caption{The difference between a given theory and the emulator results is shown for redshifts $z=0, 1, 2, 3, 4$ and $5$. The difference between the Beyond Zel'dovich approximation and the emulator is shown in solid lines in all four panels. The differences between the Zel'dovich approximation (upper-left panel), SPT 1-loop (upper-right panel), LPT 1-loop (lower-left panel) and 3-point CLPT (lower-right panel) are shown in dashed lines. The grey shaded region shows $\Delta_\mathrm{diff}\pm{0.05}$.} 
 \label{fig:power compare}
\end{figure*}

In this section, we will compare the Beyond Zel'dovich approximation (with $\bzepsilon=1$ and $k_c=6\ \mathrm{h}\ \mathrm{Mpc}^{-1}$) power spectrum calculated using the \textsc{CTM Module} to other methods. The first method we compared the Beyond Zel'dovich approximation to is the Zel'dovich approximation which has also been computed with an initial Gaussian damped power spectrum. In Figure~\ref{fig:power compare}, the top-left panel shows the difference between the Beyond Zel'dovich (solid lines) and Zel'dovich (dashed) power spectra at different redshifts. Above $z=1$ the Beyond Zel'dovich approximation matches the \textsc{Euclid Emulator} results more consistently. 

In the upper-right hand panel of Figure~\ref{fig:power compare}, the difference between the Beyond Zel'dovich approximation and SPT 1-loop (dashed lines) is shown for a range of redshifts. The SPT 1-loop power spectra and correlation function were calculated using \textsc{fastpt}~\citep{McEwen2016, Fang2017}. SPT 1-loop models the non-linear regime more accurately than the Beyond Zel'dovich approximation at all redshifts. However, in Section~\ref{subsec: compare bz corr func} we observed that SPT 1-loop does not model the BAO feature in the correlation function as accurately as other methods. 

The difference between the Beyond Zel'dovich approximation power spectrum, that obtained for LPT 1-loop and the emulator is shown in the lower left panel of Figure~\ref{fig:power compare}. At redshifts less than $z=4$, LPT 1-loop models structure formation on small scales more accurately. For redshifts $z=4$ and $z=5$, the Beyond Zel'dovich approximation performs as well LPT 1-loop until around $k=0.1\ \mathrm{h}\ \mathrm{Mpc}^{-1}$.

Finally, in the bottom-right panel the Beyond Zel'dovich approximation is compared to 3-point Convolution Lagrangian Perturbation Theory (CLPT) and computed using \textsc{CLEFT}\footnote{ https://github.com/modichirag/CLEFT} (the CLPT correlation function presented in Section~\ref{subsec: compare bz corr func} was also calculated this way). Again CLPT matches the emulator results more stringently in the non-linear regime at all redshifts compared to the Beyond Zel'dovich approximation. In summary, the Beyond Zel'dovich approximation is more accurate than the Zel'dovich approximation above redshift $z=1$ and matches LPT 1-loop until around $k=0.1\ \mathrm{h}\ \mathrm{Mpc}^{-1}$. 

\section{Conclusions}

In this paper, we have introduced the Cosmological Trajectories Method (CTM). The leading second-order CTM trajectory, the Beyond Zel'dovich approximation is comprised of the Zel'dovich approximation with a gravitational correction term given by the product of the linear displacement field and a tidal tensor. This post-Born approximation to the Zel'dovich approximation should capture non-linear effects such as gravitational deflection. We then introduced a special case of second-order CTM called the Beyond Zel'dovich approximation in which the linear order terms were proportional to the linear growth factor, $\lingrowth$. We have calculated the exact expression for the Beyond Zel'dovich matter power spectrum, assuming Gaussian initial conditions.  A numerical implementation of this expands around this solution, for stability. The Beyond Zel'dovich approximation computed with an initial Gaussian damped power spectrum outperformed the Zel'dovich approximation (also computed with an initial Gaussian damped power spectrum) when compared to the power spectrum obtained using the \textsc{Euclid Emulator} at redshifts above $z=1$. The Beyond Zel'dovich approximation also matched the emulator correlation function between $r=5\ \mathrm{Mpc}\ \mathrm{h}^{-1}$ and $r=10\ \mathrm{Mpc}\ \mathrm{h}^{-1}$ as well as SPT 1-loop, LPT 1-loop and CLPT at $z=0$. As demonstrated in Figure~\ref{fig:corr percent diff} the Beyond Zel'dovich approximation further models the BAO peak in the correlation function more accurately than the Zel'dovich approximation, SPT 1-loop and CLPT at $z=0$.

The Beyond Zel'dovich approximation power spectrum also matched the performance of LPT 1-loop until $k=0.1\ \mathrm{h}\ \mathrm{Mpc}^{-1}$ when compared to the emulator power spectrum above $z=1$. Although LPT 1-loop, SPT 1-loop and CLPT outperformed the Beyond Zel'dovich approximation on small scales ($k\geq0.5\ \mathrm{h}\ \mathrm{Mpc}^{-1}$), on mildly non-linear scales the Beyond Zel'dovich approximation matched the emulator power spectrum. This suggests that the CTM could be implemented to produce mock observables for future BAO observations taken by instruments such as \textit{DESI} in its current state. 

We investigated and fixed the two free parameters of the theory, $\bzepsilon$ which controls the size of the correction term, and $z_i$ the initial redshift at which the correlations of the fields are calculated at to be $\bzepsilon=1$ and $z_i=100$. On small scales ($k\geq{1}\ \mathrm{h}\ \mathrm{Mpc}^{-1}$) numerical integration issues were encountered, resulting in the distrust of results beyond this point. The numerical issues and possible solutions were discussed in Appendix~\ref{appen ni: comments on ni}. We introduced an initial Gaussian damped power spectrum with a cut-off scale of $k_c=6\ \mathrm{h}\ \mathrm{Mpc}^{-1}$ to reduce the excessive damping on mildly non-linear scales due to shell-crossing showed in Figure~\ref{fig:breakdown}. We leave it to future work to investigate the dependence of this Gaussian cut-off on both cosmology and redshift as it may have an impact on the application of the CTM to modified gravity theories or large deviations from $\Lambda$-CDM cosmology.

In conclusion, both the CTM and the Beyond Zel'dovich approximation appear to be valuable tools for studying nonlinear clustering of matter and galaxies in the Universe. The Beyond Zel'dovich approximation can be used to 
interpret future BAO observations by instruments such as \textit{DESI} and \textit{LSST}. Furthermore, both approximations could be used in conjunction with Lyman-$\alpha$ observations, re-ionisation studies and other high redshift surveys, as we verified that our approach performs particularly well at higher redshift. Thus, the CTM promises to enable placing even tighter statistical constraints on viable models of dark matter and dark energy, as well as on modified gravity theories.

\section*{Acknowledgements}

The authors would like to thank Matthias Bartelmann, Yacine Ali-Haimoud, Zvonimir Vlah and Yanchuan Cai for useful discussions. F.C. Lane acknowledges the support of the UK Science and Technology Facilities Council and the Scottish Universities Physics Alliance. A.N. Taylor thanks the Royal Society for the support of a Wolfson Research Merit Award and a STFC Consolidated Grant. D. Sorini is supported by the European Research Council, under grant no. 670193.



\bibliographystyle{mnras}
\bibliography{example} 



\appendix

\section{The CTM trajectory calculation}
\label{appen gtc: general traj calc}

The CTM trajectory is an expansion of the gravitationally induced trajectory around the Zel'dovich approximation, 
given by

\begin{equation}
    \label{appen gtc: general traj}
    \mathbf{x}\left(\mathbf{q},t\right)=\mathbf{q}+A\left(t\right)\mathbf{\Psi}\left(\mathbf{q},t_i\right)
    +
    \bzepsilon\int_{t_i}^{t} \!\!
    \frac{dt'}{a'^2}\int_{t_i}^{t'}
    \! \! dt'' \Delta \mathbf{g} 
    \left(\mathbf{x}\left(\mathbf{q},t''\right),t''\right)
\end{equation}

\noindent where $a$ is the scale factor, $\mathbf{q}$ is the initial position, $\mathbf{\Psi}$ is the linear order displacement field,  $t_i$ is some initial time. The gravitational field is given by $ \mathbf{g} = - \nabla_{\mathbf{x}}\phi$ where $\phi$ is the gravitational potential and $\bzepsilon$ is an expansion parameter used to control the size of the higher-order gravitational term. As the Zel'dovich approximation already extrapolates the effects of the linear gravitational field, we add the differential gravitational field $\Delta \mathbf{g} = \mathbf{g}- \mathbf{g}_\mathrm{L}$, where we have removed the linear field to avoid double-counting forces. The CTM trajectory in Equation~(\ref{appen gtc: general traj})  then describes a particle moving under free motion given by the Zel'dovich approximation with the addition of a gravitational correction term.

The CTM approach is a hybrid of KFT~\citep{bartelmann2014, bartelmann2014a, alihaimoud2015} and LPT~\citep{Moutarde1991,catelan1995,Buchert1992a,Buchert1993,bouchet1996,tatekawa2004,Rampf2012}. In the KFT approach, the trajectory is based on the formal solution to the particle equations of motion, and so have the free particle motion is damped by the expansion. Here we have chosen to have the trajectory in Equation~(\ref{appen gtc: general traj}) to be defined by the Zel'dovich approximation plus a higher-order gravitational term we are free to pick the time-dependent function, $A\left(t\right)$, to be the linear growth factor, $\lingrowth$, which will allow us to avoid re-normalisation on large scales. In the remainder of this section, we will demonstrate how the second-order CTM trajectory, given in Equation~(\ref{eqn: general trajectory full tidal}) is obtained by solving the gravitational field using the Zel'dovich approximation as a basis. This allows us to avoid the inclusion of non-local terms, which arise when considering second-order LPT.

We can write the trajectory~(\ref{appen gtc: general traj}) as 
\begin{equation}
    \label{appen gtc: general traj split up}
    \mathbf{x}\left(\mathbf{q},t\right)=\mathbf{q}+\mathbf{x}_0\left(\mathbf{q},t\right)+\mathbf{x}_1\left(\mathbf{q},t\right)
\end{equation}
\noindent with

\begin{align}
    \label{appen gtc: general traj split up:1}
    \mathbf{x}_0\left(\mathbf{q},t\right)&=A\left(t\right)\mathbf{\Psi}\left(\mathbf{q},t_i\right),
    \\
    \mathbf{x}_1\left(\mathbf{q},t\right)&=\bzepsilon\int_{t_i}^{t}\frac{dt'}{a'^2}\int_{t_i}^{t'}dt'' \Delta \mathbf{g}\left(\mathbf{x}\left(\mathbf{q},t''\right),t''\right)
\end{align}

\noindent The overdensity field, $\delta\left(\mathbf{x},t\right)$, is given by
\begin{equation}
    \label{appen gtc: overden field}
    \delta\left(\mathbf{x},t\right)=\int{d^3q\ }\diracdelta\left(\mathbf{x}-\mathbf{q}-\mathbf{x}_0-\mathbf{x}_1\right)-1
\end{equation}
\noindent and its Fourier transform \footnote[1]{Our Fourier transform convention is $f\left(\mathbf{k}\right)=\int{d^3x\ }\mathrm{e}^{i\mathbf{k}\cdot\mathbf{x}}f\left(\mathbf{x}\right)$ and $f\left(\mathbf{x}\right)=\frac{1}{\left(2\pi\right)^3}\int{d^3k\ }\mathrm{e}^{-i\mathbf{k}\cdot\mathbf{x}}f\left(\mathbf{k}\right)$.} is
\begin{equation}
    \label{appen gtc: overden field ft}
    \left(2\pi\right)^3\diracdelta\left(\mathbf{k}\right)+\delta\left(\mathbf{k},t\right)= \int{d^3q\ }\mathrm{e}^{i\mathbf{k}\cdot\mathbf{q}}\mathrm{e}^{i\mathbf{k}\cdot\left(\mathbf{x}_0+\mathbf{x}_1\right)}.
\end{equation}
\noindent Using the Poisson equation, $\nabla^2\phi=\frac{3}{2}H^2_0\Omega_ma^{-1}\delta=\frac{3}{2}\omega^2_0a^{-1}\delta$, with $
\omega^2_0=H^2_0\Omega_m$ the gravitational field $\mathbf{g}\left(\mathbf{x},t\right)$ can be written as
\begin{equation}
    \label{appen gtc: grav field}
    \mathbf{g}\left(\mathbf{x},t\right)=i\omega_0^2a^{-1}\int\frac{d^3k}{\left(2\pi\right)^3}\ \mathrm{e}^{-i\mathbf{k}\cdot\mathbf{x}}\frac{\mathbf{k}}{k^2}\delta\left(\mathbf{k},t\right)
\end{equation}
\noindent where $\delta\left(\mathbf{k},t\right)$ is the  non-linear overdensity field. The overdensity field is given  by the Zel'dovich approximation,
\begin{equation}
    \label{appen gtc: linear overden}
    \delta\left(\mathbf{x},t\right)=\int{d^3q\ }\diracdelta\left(\mathbf{x}-\mathbf{q}-\mathbf{x}_0\right)-1,
\end{equation}
and the linear displacement is given by
\begin{equation}
    \mathbf{\Psi} = \int\frac{d^3k}{\left(2\pi\right)^3}\ \mathrm{e}^{-i\mathbf{k}\cdot\mathbf{x}}\frac{\mathbf{k}}{k^2}\delta^{(0)}\left(\mathbf{k},t\right).
\end{equation}

\noindent These equations define our Cosmological Trajectories Method.

The Beyond Zel'dovich approximation calculates the gravitational correction to second order in the displacement field. To this order the gravitational field is 

\begin{equation}
    \mathbf{g} = \omega_0^2 \frac{1}{a} A(t) \left(\mathbf{\Psi} 
    - A(t)\left[(\mathbf{\Psi} \cdot \boldsymbol{\nabla}) \mathbf{\Psi} 
    + \boldsymbol{\nabla} \nabla^{-2} \left(\bar E^2 - |\delta^{(0)}|^2 \right)\right]\right),
    \label{eq:grav}
\end{equation}

\noindent where $\bar E^2 = \bar E_{ij} \bar E_{ji}$, the tidal tensor is $\bar{E}_{ij}=\nabla_i\nabla_j\nabla^{-2}\delta^{\left(0\right)}$ and $\delta^{\left(0\right)}$ is the linear overdensity field. The first term in equation (\ref{eq:grav}) is the linear gravitation field, while the second term represents a local change from Lagrangian to Eulerian coordinates. The last term is the force generated by nonlinear, second-order growth of structure. This term is non-local, depending on all points in the density and tidal field through the inverse Laplacian. However, we can keep our analysis local as we can expect the second term, proportional to the displacement field which we will extrapolate, to be larger than the third term. In addition, we can expect the cancellation between $\bar E^2$ and $|\delta^{(0)}|^2$ in the third terms to reduce its effects. Hence the differential gravitational field, to leading second-order, is 

\begin{align}
    \label{appen gtc: expanded grav field}
 \Delta \mathbf{g}&\approx-\omega_0^2\frac{1}{a}A\left(t\right)^2\left[\mathbf{\Psi}\left(\mathbf{q},t_i\right)\cdot \boldsymbol{\nabla}\right] \mathbf{\Psi}\left(\mathbf{q},t_i\right).
\end{align}

\noindent Therefore, we define the Beyond Zel'dovich approximation as

\begin{equation}
    \label{appen gtc: final general traj}
    {x}_i\left(\mathbf{q},t\right)={q}_i+\Psi_i\left(\mathbf{q},t_i\right)\left[A\left(t\right)\delta_{ij}+B_\epsilon\left(t\right)\bar{E}_{ij}\left(\mathbf{q},t_i\right)\right],
\end{equation}
\noindent  where the tidal term describes the effects of gravitational scattering.

The time-dependent function $B_\epsilon\left(t\right)$ is, 

\begin{equation}
    \label{appen gtc: B in terms of t}
    B_\epsilon\left(t\right)=-\bzepsilon\omega_0^2\int_{t_i}^{t}\frac{dt'}{a'^2}\int_{t_i}^{t'}\frac{dt''}{a''}A\left(t''\right)^2,
\end{equation}

\noindent which after using $dt=-\frac{a}{H}dz$ can be written as, 

\begin{equation}
    \label{appen gtc: B}
    B_\epsilon\left(t\right)=-\bzepsilon\omega_0^2\int_{z_i}^{z}\frac{dz'}{a'H\left(z'\right)}\int_{z_i}^{z'}\frac{dz''}{H\left(z''\right)}A\left(z''\right)^2.
\end{equation}

\noindent The Beyond Zel'dovich approximation used in the main body of the paper is the second-order CTM trajectory given in Equation~(\ref{appen gtc: final general traj}) with the following time-dependent functions

\begin{align}
    \label{appen gtc: time dep funcs BZ}
    A\left(z\right)&=\frac{\lingrowth\left(z\right)}{\lingrowth\left(z_i\right)},
    \\
    B_\epsilon\left(z\right)&=-\bzepsilon\omega_0^2\int_{z_i}^{z}\frac{dz'}{a'H\left(z'\right)}\int_{z_i}^{z'}\frac{dz''}{H\left(z''\right)}\left(\frac{\lingrowth\left(z''\right)}{\lingrowth\left(z_i\right)}\right)^2.
\end{align}

\noindent This time dependence reproduces linear growth  on large scales without the need for re-normalisation.

\section{Correlation functions}
\label{appen: corr funcs}

In Section~\ref{subsubsec: covariance matrix} we defined the covariance matrix, $\mathbf{C}=\langle\mathbf{X}\mathbf{X}^T\rangle$, where $X_\alpha=\left(\Psi_i^1, \Psi_i^2, E^1_{i}, E^2_{i}, \delta^{\left(0\right)}_1, \delta^{\left(0\right)}_2\right)$.  In order to evaluate the CTM power spectrum it is useful to have the correlation functions defined in Equations~(\ref{eqn: defs of correlations: sigma_ij}, \ref{eqn: defs of correlations: phi_ijk},  \ref{eqn: defs of correlations: eta_ijkl}, \ref{eqn: defs of correlations: Sigma_ij}, \ref{eqn: defs of correlations: pi_i}, \ref{eqn: defs of correlations: sigma_0}) re-expressed in of $\bar{E}_{ij}$. For example, 
\begin{align}
\notag
{\Phi}_{ijk}\left(q\right)&=\langle{E}_{ij}\left(\mathbf{q}_1\right)\Psi_k\left(\mathbf{q}_2\right)\rangle
\\
\notag
&=\left\langle\left(\bar{E}_{ij}\left(\mathbf{q}_1\right)-\frac{1}{3}\delta^{\left(0\right)}\left(\mathbf{q}_1\right)\right)\Psi_k\left(\mathbf{q}_2\right)\right\rangle
\\
\label{eqn: phi_ijk}
&=\bar{\Phi}_{ijk}\left(\mathbf{q}\right)-\frac{1}{3}\Pi_k\left(\mathbf{q}\right)\delta_{ij}
\end{align}
 
\noindent where 
\begin{subequations}
\label{eqn: covariance defs one}
\begin{align}
    \label{eqn: covariance defs: phi}
    \bar{\Phi}_{ijk}\left(q\right)&=\langle\bar{E}_{ij}\left(\mathbf{q}_1\right)\Psi_k\left(\mathbf{q}_2\right)\rangle=i\int\frac{d^3k}{\left(2\pi\right)^3}\ \mathrm{e}^{i\mathbf{k}\cdot\mathbf{q}}\frac{k_ik_jk_k}{k^4}\linearpow\left(k,z_i\right),
    \\
    \label{eqn: covariance defs: pi}
    \Pi_i\left(q\right)&=\langle{\Psi}_{i}\left(\mathbf{q}_1\right)\delta^{\left(0\right)}\left(\mathbf{q}_2\right)\rangle=i\int\frac{d^3k}{\left(2\pi\right)^3}\ \mathrm{e}^{i\mathbf{k}\cdot\mathbf{q}}\frac{k_i}{k^2}\linearpow\left(k,z_i\right),
\end{align}
\end{subequations}
\noindent with $\linearpow\left(k,\ z_i\right)$ being the linear power spectrum evaluated at the initial redshift. Similarly,

\begin{align}
    \label{eqn: eta_ijkl}
    \eta_{ijkl}\left(q\right)&=\bar{\eta}_{ijkl}\left(q\right)+\frac{1}{9}\xi_0\left(q\right)\delta_{ij}\delta_{kl}-\frac{1}{3}\left(\bar{\Sigma}_{ij}\left(q\right)\delta_{kl}+\bar{\Sigma}_{kl}\left(q\right)\delta_{ij}\right),
    \\
    \label{eqn: sigma_in}
    \Sigma_{ij}\left(q\right)&=\bar{\Sigma}_{ij}\left(q\right)-\frac{1}{3}\xi_0\left(q\right)\delta_{ij},
\end{align}

\noindent with 
\begin{subequations}
\label{eqn: covariance defs two}
\begin{align}
    \label{eqn: covariance defs two: eta}
    \bar{\eta}_{ijkl}\left(q\right)&=\langle\bar{E}_{ij}\left(\mathbf{q}_1\right)\bar{E}_{kl}\left(\mathbf{q}_2\right)\rangle=\int\frac{d^3k}{\left(2\pi\right)^3}\ \mathrm{e}^{i\mathbf{k}\cdot\mathbf{q}}\frac{k_ik_jk_kk_l}{k^4}\linearpow\left(k,z_i\right),
    \\
    \label{eqn: covariance defs two: Sigma}
    \bar{\Sigma}_{ij}\left(q\right)&=\langle\bar{E}_{ij}\left(\mathbf{q}_1\right)\delta^{\left(0\right)}\left(\mathbf{q}_2\right)\rangle=\int\frac{d^3k}{\left(2\pi\right)^3}\ \mathrm{e}^{i\mathbf{k}\cdot\mathbf{q}}\frac{k_ik_j}{k^2}\linearpow\left(k,z_i\right),
    \\
    \label{eqn: covariance defs two: sigma_0}
    \xi_0\left(q\right)&=\langle\delta^{\left(0\right)}\left(\mathbf{q}_1\right)\delta^{\left(0\right)}\left(\mathbf{q}_2\right)\rangle=\int\frac{d^3k}{\left(2\pi\right)^3}\ \mathrm{e}^{i\mathbf{k}\cdot\mathbf{q}}\ \linearpow\left(k,z_i\right).
\end{align}
\end{subequations}

\subsection{An Example of the splitting of a correlation function}
\label{appen: corr funcs example}

We will implement the identities presented in~\citet{catelan2000} and~\citet{crittenden2001} to calculate $\bar{\Sigma}_{ij}$ and $\Pi_{i}$. For example, $\bar{\Sigma}_{ijk}$ given in Equation~(\ref{eqn: covariance defs two: Sigma}) can be written as 

\begin{equation}
    \label{appen eqn: sigma alignment method 1}
    \bar{\Sigma}_{ij}\left(q,z_i\right)=\frac{1}{\left(2\pi\right)^2}\int_{-1}^{1}{d\mu\ }\mathrm{e}^{ikq\mu}\int_0^{\infty}dk\ k_ik_j\linearpow\left(k,z_i\right)
\end{equation}

\noindent which after performing the angle integral and substituting in $ik_i=\nabla_i$ results in 

\begin{equation}
    \label{appen eqn: sigma alignment method}
    \bar{\Sigma}_{ij}\left(q,z_i\right)=-\frac{1}{2\pi^2}\int_0^{\infty}dk\ \linearpow\left(k, z_i\right)\nabla_i\nabla_jj_0\left(kq\right).
\end{equation}

\noindent Defining the following

\begin{equation}
    \nabla_i=q_i\frac{1}{q}\frac{d}{dq}=q_iD_q\ \mathrm{and\ }D^n_rj_0\left(r\right)=\left(-1\right)^nr^{-n}j_n\left(r\right)
\end{equation}

\noindent where $r=kq$ in our case allows us to decompose $\bar{\Sigma}_{ij}$ as

\begin{equation}
    \label{appen eqn: decomposed sigma}
    \bar{\Sigma}_{ij}\left(q,z_i\right)=D\left(q,z_i\right)\delta_{ij}+F\left(q, z_i\right)\hat{q}_i\hat{q}_j
\end{equation}

\noindent with $D\left(q,z_i\right)$ and $F\left(q,z_i\right)$ as defined in Equations~(\ref{eqn: D, F and G definitions: D definition}) and~(\ref{eqn: D, F and G definitions: F definition}).

\section{Comments on Numerical Integration}
\label{appen ni: comments on ni}

In Section~\ref{sec: the BZ approx} it is mentioned that we only trust the CTM power spectrum up until $k=0.9\ \mathrm{h}\ \mathrm{Mpc}^{-1}$. After this point, there are numerical uncertainties due to the highly oscillatory spherical Bessel integrals involved in the calculation. We investigated multiple techniques to remedy these numerical issues for large-$k$ values. 

We first implemented an alternative numerical integration technique to the one introduced in Section~\ref{subsubsec: calc expanded power spec}. This alternate technique was introduced in \citet{vlah2015a} and involves a generalisation of the plane wave expansion~\citep{planewaves}. We will briefly summarise this alternative integration technique here but refer the reader to \citet{vlah2015a} for a full derivation. The plane wave expansion is 

\begin{equation}
\label{appen eqn: plane wave expansion}
\mathrm{e}^{i\mathbf{k}\cdot\mathbf{q}}=\sum_{l=0}^{\infty}i^l\left(2l+1\right)\legendre\left(\cos\theta\right)j_l\left(kr\right)
\end{equation}

\noindent using this we can write that

\begin{equation}
\label{appen eqn: expansion of plane wave}
\left(ix\right)^n=\sum_{l=0}^{\infty}i^l\left(2l+1\right)\legendre\left(x\right)\left(\frac{d^nj_l\left(\alpha\right)}{d\alpha^n}\right)_{\alpha=0}.
\end{equation}

\noindent This is simply the Taylor expansion of the spherical Bessel function around zero. Comparing this Taylor expansion with another well known representation of spherical Bessel function~\footnote{https://dlmf.nist.gov}, 

\begin{equation}
\label{appen eqn: series rep of spherical bessel}
j_l\left(\alpha\right)=\alpha^l\sum_{k=0}^{\infty}\frac{\left(-1\right)^k}{2^kk!}\frac{\alpha^{2k}}{\left(2l+2k+1\right)!!},
\end{equation}

\noindent we can write Equation~(\ref{appen eqn: plane wave expansion}) as

\begin{equation}
\label{appen eqn: expansion of plane wave 1}
\left(ix\right)^n=\sum_{l=0}^{\infty}i^l\left(2l+1\right)\legendre\left(x\right)b^l_n
\end{equation}

\noindent where

\begin{equation}
\label{appen eqn: b defintion}
  b^l_n=\left\{
  \begin{array}{cc}
    \frac{i^{n-l}n!}{\sqrt{2}^{n-l}\left(\frac{1}{2}\left(n-l\right)\right)!,\left(n+l+1\right)!!}, & \text{if}\ n\geq{l}\ \text{and\ } n\ \text{and\ }l\ \text{are\ both\ even\ or\ odd}\\
    0 , & \text{otherwise.}
  \end{array}\right.
\end{equation} 

\noindent Therefore,

\begin{multline}
\label{appen eqn: alt integral identity}
\int_{-1}^{1}{d\mu\ }\mathrm{e}^{iA\mu}\mathrm{e}^{B\mu^2}=2\sum_{n=0}^{\infty}\frac{\left(2n\right)!}{2^nn!}B^n
\\
\times\sum_{p=0}^n\left(-2\right)^p\frac{4p+1}{\left(n-p\right)!\left(2n+2p+1\right)!!}j_{2p}\left(A\right).
\end{multline}

\noindent In~\citet{vlah2015a} this integration technique was used to calculate both the LPT 1-loop and CLPT power spectra. It was found that there was only a difference between this method and the method used in this paper for high $k$-values. We also reached the same conclusion in regards to the CTM power spectrum. The method in Section~\ref{subsubsec: calc expanded power spec} has numerical advantages as it contains only one infinite sum, hence it was this method that we implemented in the \textsc{CTM Module}.

In order to calculate the infinite sum numerically in Equation~(\ref{eqn: final traj power spec}) we truncate the sums at $n=32$. To reduce the impact of the higher-order spherical Bessel functions on the summation we investigated the impact of truncating the sums at $n=10$ instead. We found that this removed some of the numerical noise, however, did not impact the maximum $k$-value reached before we dropped below $5\%$ of the \textsc{Euclid Emulator}.

The spherical Bessel functions in this paper have been calculated using the publicly available \textsc{mcfit}. This software is based on the FFTLog algorithm~\citep{fftlog} and the \textsc{FFTLog} code~\footnote{https://jila.colorado.edu/~ajsh/FFTLog/}. Although these codes can be fully optimised to calculate the Zel'dovich power spectrum, we encountered issues when the correction term in the CTM trajectory becomes large for either large $k$-values or low redshifts. We leave it to future work to implement an original integration routine, fully optimised for the CTM power spectrum.

\section{Application of the CTM to KFT}
\label{appen gtc: application to kft}

In \citet{alihaimoud2015} a more detailed computation of the power spectrum to first-order in the gravitational interaction is given. The computation is also described in \citet{bartelmann2014, bartelmann2014a} from the statistical mechanics perspective. We will simply summarise the results here so that we may compare our power spectrum to that presented in  \citet{bartelmann2014} and \citet{bartelmann2014a}. Using the definition of the overdensity field~(\ref{appen gtc: overden field}) the Dirac delta can be expanded such that

\begin{subequations}
    \label{appen eqn: split overden ah}
    \begin{equation}
        1+\delta^{\left(0\right)}\left(\mathbf{x}\right)=\int\diracdelta\left(\mathbf{x}-\mathbf{x}_0\right){d^3q},
    \end{equation}
    \begin{equation}
        \delta^{\left(1\right)}\left(\mathbf{x}\right)=-\int\mathbf{x}_1\left(\mathbf{q}\right)\cdot\nabla\diracdelta\left(\mathbf{x}-\mathbf{x}_0\right){d^3q}.
    \end{equation}
\end{subequations}

\noindent To first-order in the gravitational interaction (with $\mathrm{P}^{\left(00\right)}\propto\langle\delta^{\left(0\right)}\delta^{\left(0\right)}\rangle$ and $\mathrm{P}^{\left(01\right)}\propto\langle\delta^{\left(0\right)}\delta^{\left(1\right)}\rangle$) the power spectrum is given by,

\begin{equation}
    \label{appen eqn: power spec expan ah}
    \mathrm{P}\left(k\right)\approx\mathrm{P}^{\left(00\right)}\left(k\right)+2\epsilon\mathrm{P}^{\left(01\right)}\left(k\right).
\end{equation}

\noindent It is then noted that $\mathbf{x}_0$ is equal to the Zel'dovich approximation with the exception of the time dependent function $\alpha\left(t\right)=1+a_i^2\frac{\dot{D}_1\left(t_i\right)}{\lingrowth\left(t_i\right)}\int^t_{t_i}\frac{dt'}{a'^2}$. Therefore, the power spectrum to zeroth order in the interaction is,

\begin{equation}
    \label{appen eqn: p00}
    \mathrm{P}^{\left(00\right)}\left(k\right)=\int{d^3q}\ \mathrm{e}^{i\mathbf{k}\cdot\mathbf{q}}\left[\mathrm{e}^{-k^2\alpha^2\sigma^2_{\psi}+k_ik_j\alpha^2\sigma_{ij}\left(\mathbf{q}\right)}-1\right].
    \end{equation}

\noindent Calculating $\mathrm{P}^{\left(01\right)}$ is more involved as it requires taking the correlation of $\delta^{\left(0\right)}$ and $\delta^{\left(1\right)}$. It is calculated in full in \citet{alihaimoud2015}, \citet{bartelmann2014} and \citet{bartelmann2014a}. However, let us focus on the result obtained if one expands in terms of the linear power spectrum

\begin{subequations}
    \label{appen eqn: ah expansion in P_L}
    \begin{equation}
        \mathrm{P}^{\left(00\right)}_\mathrm{L}\left(k,z\right)=\alpha^2\left(z\right)\linearpow\left(k,z_*\right),
    \end{equation}
    \begin{equation}
        \mathrm{P}^{\left(01\right)}_\mathrm{L}\left(k,z\right)=\omega_0^2\alpha\left(z\right)\int^{t}_{t_*}\frac{dt'}{a'^2}\int_{t_*}^{t'}\frac{dt''}{a''}\alpha\left(z''\right)\linearpow\left(k,z_*\right),
    \end{equation}
\end{subequations}

\noindent where $\omega_0^2=\frac{3}{2}H_0^2\Omega_m$. There are a number of issues raised in \citet{alihaimoud2015} concerning the results presented in \citet{bartelmann2014} and \citet{bartelmann2014a}. One such issue is with the expansion carried out to calculate the power spectrum. Expanding the Zel'dovich power spectrum in the usual way gives~\citep{crocce2006}

\begin{equation}
    \label{appen eqn: expanding zel power}
    \mathrm{P}_{\mathrm{zel}}\left(k\right)\approx\linearpow\left(k\right)-k^2\sigma^2_{\psi}\linearpow\left(k\right)+\mathrm{P}_{1-\mathrm{loop}}\left(k\right)
\end{equation}

\noindent with

\begin{equation}
    \label{appen eqn: 1 loop def}
\mathrm{P}_{1-\mathrm{loop}}\left(k\right)=\frac{1}{2}\int\frac{d^3k'}{\left(2\pi\right)^3}\frac{\linearpow\left(k'\right)\linearpow\left(k''\right)}{k'^4k''^4}\left(\mathbf{k}\cdot\mathbf{k}'\right)^2\left(\mathbf{k}\cdot\mathbf{k}''\right)^2.
\end{equation}

\noindent where $\mathbf{k}''=\mathbf{k}-\mathbf{k'}$. The following expansion however, is chosen in \citet{bartelmann2014} and \citet{bartelmann2014a}

\begin{equation}
    \label{appen eqn: bart expansion power}
    \mathrm{P}_{\mathrm{zel,B14}}\left(k\right)\approx\linearpow\left(k\right)+\frac{\mathrm{P}_{1-\mathrm{loop}}}{1+k^2\sigma^2_{\psi}}.
\end{equation}

\noindent This may lead to enhancement of power on small scales, which is what one would expect from the non-linear power spectrum. Note that there are two free parameters in this approach. There is the time at which the new trajectory is ``switched on'', $z_*$ and there is the book-keeping parameter, $\epsilon$. In \citet{alihaimoud2015} $z_*$ is chosen to be $z_*=99$ and $\epsilon=1$ to match the results in \citet{bartelmann2014} and \citet{bartelmann2014a}.

The second-order CTM time-dependent functions $A\left(z\right)$ and $B_\epsilon\left(z\right)$ defined in Equation~(\ref{appen gtc: B}) for KFT are 

\begin{subequations}
    \label{appen eqn: A and B definitions AH15}
    \begin{align}
    A\left(z\right)&=\alpha\left(z\right)+\bzepsilon\omega^2_0\beta\left(z\right),
    \\
    B_\epsilon\left(z\right)&=-\bzepsilon\omega^2_0\gamma\left(z\right).
    \end{align}
\end{subequations}

\noindent with
 
\begin{subequations}
    \label{appen eqn: time dep funcs ah15}
    \begin{equation}
        \alpha\left(z\right)=1+a_i{H\left(z_i\right)}\frac{D'_1\left(z_i\right)}{\lingrowth\left(z_i\right)}\int^z_{z_i}\frac{dz'}{a'H\left(z'\right)},
    \end{equation}
    \begin{equation}
        \beta\left(z\right)=\int^z_{z_i}\frac{dz'}{a'H\left(z'\right)}\int^{z'}_{z_i}\frac{dz''}{H\left(z''\right)}\frac{\lingrowth\left(z''\right)}{\lingrowth\left(z_i\right)},
    \end{equation}
    \begin{equation}
        \gamma\left(z\right)=\int^z_{z_i}\frac{dz'}{a'H\left(z'\right)}\int^{z'}_{z_i}\frac{dz''}{H\left(z''\right)}\frac{\lingrowth\left(z''\right)}{\lingrowth\left(z_i\right)}\alpha\left(z''\right).
    \end{equation}
\end{subequations}

\begin{figure}
 \includegraphics[width=\columnwidth]{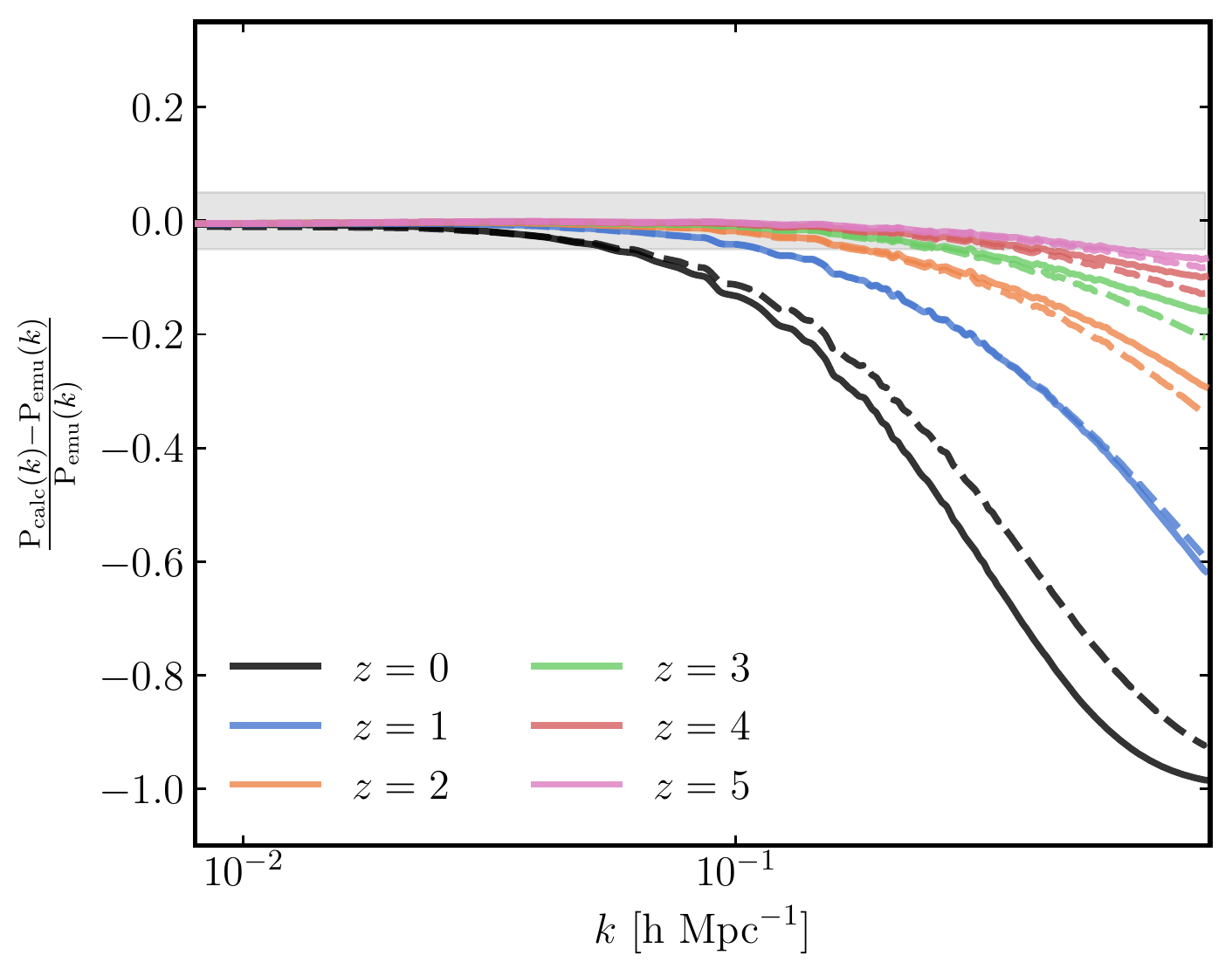}
 \caption{The difference between a given theory and the emulator results is shown for redshifts $z=0, 1, 2, 3, 4$ and $5$. The difference between the Beyond Zel'dovich approximation and the emulator is shown in solid lines and the differences between KFT and the emulator is shown in dashed lines. The grey shaded region shows $\Delta_\mathrm{diff}\pm{0.05}$.}
 \label{fig:kft comparison}
\end{figure}

\noindent The difference between the Beyond Zel'dovich approximation (solid lines) and KFT calculated using the CTM (dashed lines) and the emulator is shown in Figure~\ref{fig:kft comparison} for $z=0, 1, 2, 3, 4, 5$. Both the Beyond Zel'dovich approximation and KFT were calculated using $\bzepsilon=1$, $z_i=100$ and $k_c=6\ \mathrm{h}\ \mathrm{Mpc}^{-1}$. At redshifts, $z=2$ and above the Beyond Zel'dovich approximation outperforms KFT (calculated using the CTM) when compared to the \textsc{Euclid Emulator}. This is likely due to the time-dependent functions $A\left(z\right)$ and $B_\epsilon\left(z\right)$ being marginally larger in the Beyond Zel'dovich approximation.

Part of the motivation for the introduction of the CTM and the Beyond Zel'dovich approximation was that KFT does not regain linear growth on large scales as expected. Thus, in Figure~\ref{fig:kft comparison} the KFT results have been re-normalised by a factor of $A^{-2}\left(z\right)\left(\frac{\lingrowth\left(z_i\right)}{{\lingrowth\left(z\right)}}\right)^2$.


\bsp	
\label{lastpage}
\end{document}